\documentclass[journal]{IEEEtran}
\usepackage{cite}

\ifCLASSINFOpdf
\else
\fi
\usepackage[ruled,linesnumberedhidden]{algorithm2e}

\usepackage{amsmath,amssymb,amsfonts}
\usepackage[pdftex]{graphicx}

\makeatletter
\newcommand{\rmnum}[1]{\romannumeral #1}
\newcommand{\Rmnum}[1]{\expandafter\@slowromancap\romannumeral #1@}
\makeatother
\usepackage{algorithmicx} 
\usepackage{algpseudocode}  

\usepackage{color}

\usepackage{amsmath,amssymb,amsfonts}
\usepackage{array}

\usepackage{stfloats}
\begin{document}
%
\title{Robust Secure Transmission Design for IRS-Assisted mmWave Cognitive Radio Networks}
%
%
%

\author{Xuewen Wu,~\IEEEmembership{Graduate Student Member,~IEEE}, Jingxiao Ma, Chenwei Gu,  Xiaoping Xue,~\IEEEmembership{Member,~IEEE} and Xin Zeng,~\IEEEmembership{Member,~IEEE}
	\thanks{This work was supported by National Natural Science Foundation of China under Grant 61871290, Shanghai Science and Technology Innovation Action Plan under Grant 21DZ1200702. (\textit{Corresponding author: Jingxiao Ma.}) }
	\thanks{Copyright (c) 2015 IEEE. Personal use of this material is permitted. However, permission to use this material for any other purposes must be obtained from the IEEE by sending a request to pubs-permissions@ieee.org.}
	
	\thanks{{X. Wu, J. Ma, X. Xue, and X. Zeng are with the College of Electronic and Information Engineering, Tongji University, Shanghai 201804, China. (email: wuxuewen1995@163.com;  mjxiao@tongji.edu.cn; xuexp@tongji.edu.cn; zengxin1@tongji.edu.cn).}}
	\thanks{{C. Gu is with the College of Telecommunications and Information Engineering, Nanjing University of Posts and Telecommunications, Nanjing 210003, China. (email: 15706290291@139.com).}}}

%
%

\markboth{Journal of \LaTeX\ Class Files,~Vol.~14, No.~8, September~2021}%
{Shell \MakeLowercase{\textit{et al.}}: Bare Demo of IEEEtran.cls for IEEE Journals}
%



\maketitle

\begin{abstract}
	Cognitive radio networks (CRNs) and millimeter wave (mmWave) communications are two major technologies to enhance the spectrum efficiency (SE). Considering that the SE improvement in the CRNs is limited due to the interference temperature imposed on the primary user (PU), and the severe path loss and high directivity in mmWave communications make it vulnerable to blockage events, we introduce an intelligent reflecting surface (IRS) into mmWave CRNs. Due to the estimation mismatch and the passivity of Eavesdroppers (Eves), perfect channel state information (CSI) of wiretap links is challenging to obtain, which promotes our research on robust secure beamforming (BF) design in the IRS-assisted mmWave CRNs. This paper considers the collaborate scenario of Eves, which allows us to investigate the BF design in the harsh eavesdropping environment. Specifically, by using a uniform linear array (ULA) at the cognitive base station (CBS) and a uniform planar array (UPA) at the IRS, and supposing that imperfect CSIs of angle-of-departures for wiretap links are known, we formulate a constrained problem to maximize the worst-case achievable secrecy rate (ASR) of the secondary user (SU) by jointly designing the transmit BF at the CBS and reflect BF at the IRS. To solve the non-convex problem with coupled variables, an efficient alternating optimization algorithm is proposed. As for the transmit BF at the CBS, we propose a heuristic robust transmit BF algorithm to attain the BF vectors analytically. As for the reflect BF at the IRS, by means of an auxiliary variable, we transform the non-convex problem into a semi-definite programming problem with rank-1 constraint, which is handled with the help of an iterative penalty function, and then obtain the optimal reflect BF through CVX. Finally, simulation results indicate that the ASR performance of our proposed algorithm has a small gap with that of the optimal solution with perfect CSI compared with the other benchmarks.
\end{abstract}


%
\IEEEpeerreviewmaketitle

\section{Introduction}
%
%
%
%
\IEEEPARstart{W}{ith} the increasing popularity of smart devices, the data traffic of mobile communications has grown explosively in recent years \cite{1}, resulting in an impending spectrum scarcity. Cognitive radio network (CRN) and millimeter wave (mmwave) communications are two main technologies to improve spectrum efficiency (SE) \cite{2}–\cite{5}. The first one allows secondary users (SUs) employ the spectrum authorized to primary users (PUs) under the premise that the quality of service (QoS) of PUs is ensured \cite{2}. The other can supply large available bandwidth at mmWave frequencies \cite{5}.

However, there exists some problems in both CRN and mmWave communications. As for CRN, one problem is that the SE of SUs is limited by the interference temperature (IT) constraint imposed on PUs, meaning that there is a conflict between the performance improvement of PU and SU \cite{6}–\cite{9}. Specifically, increasing the transmit power at the cognitive base station (CBS) to strengthen the signal received at the SU will also bring more interference to the PU. As for mmWave communications, the signal path loss is much more severe than that over lower frequency bands \cite{22}, and high directivity makes it susceptible to blockage, especially common in dense urban environments. Besides, there exists security problems both in CRN and mmWave communications. Compared with the traditional wireless networks, security issue in the CRNs becomes more complex since SU is allowed to share spectrum with PU \cite{13}–\cite{16}. Users within the coverage area of SU’s transmitter can eavesdrop the confidential information. Due to the blockage characteristic in mmWave communications, eavesdroppers (Eves) may also block legitimate communication in addition to eavesdropping on legitimate communication, which will make the secrecy performance degrade.

To deal with the SE problem in the CRNs, many power allocation and beamforming (BF) approaches have been proposed to support the optimal transmission. Besides, in order to satisfy the interference limitation constraint, it is another approach to make the SU signals aloof from the PU by adopting multiple antennas. However, the performance improvement of these approaches is limited when the direct link from the CBS to SU is unavailable or weak. Deploying active relays or other auxiliary helpers will lead to high hardware cost and additional energy consumption. To compensate for the severe path loss and extend the coverage in mmWave communications, large antenna arrays and active relays are usually adopted for data transmission \cite{23}–\cite{25}. To deal with the security problem, some physical layer security (PLS) technologies can be utilized to ensure the secure transmission. The key point of evaluating the security performance of PLS lies in that when the transmission rate of the legitimate link is greater than that of the wiretap link, high secrecy rate (SR), which is defined by the rate difference between the legitimate link and the wiretap link, can be achieved to ensure good security \cite{17}. In order to further promote secure transmission, some technologies have been put forward and combined with PLS to reduce the security risks of eavesdropping, such as cooperative relaying \cite{18}, BF \cite{19}, zero-forcing-based BF \cite{20}, and artificial-noise (AN) injection \cite{21}. Nevertheless, active relays or other auxiliary helpers will lead to expensive hardware costs and extra power consumption. In addition, the harsh transmission environment may make it difficult to achieve satisfactory secrecy performance even with the help of AN or jamming signals.

Recently, intelligent reflecting surface (IRS) has been introduced in wireless communication networks to solve the problems mentioned above. Due to full-duplex transmission and low power consumption, IRS has received significant attention from both academia and industry as a promising technology  \cite{classic1}-\cite{classic3} to significantly increase the SE and PLS of the CRNs, and to improve the network coverage and PLS of mmWave communication systems \cite{26}-\cite{add4}. It is a new cost-effective technology that is able to change the radio propagation channels. By properly adjusting the phase shifts, some certain performance criterions such as transmit power \cite{33}-\cite{34}, transmission rate \cite{35}–\cite{37}, SR \cite{38}, secrecy energy efficiency (SEE) \cite{39} in the IRS-assisted CRNs are effectively optimized, and coverage of mmWave signals \cite{40}-\cite{add2}, received signal power \cite{40}, ergodic capacity \cite{41}, weighed sum-rate \cite{add2},\cite{add1}, SR \cite{42}-\cite{43} in the IRS-assisted mmWave communication systems are enhanced. Besides the transmission design, the channel estimation of IRS is also one of the research contents of IRS. The authors of \cite{estimation1} proposed an innovative three-phase framework to estimate a large number of channel coefficients in the IRS-assisted uplink multiuser communications accurately using merely a small number of pilot symbols. Two efficient uplink channel estimation schemes for different channel setups in the IRS-assisted multi-user OFDMA system were proposed in \cite{estimation2}. Moreover, to reduce the high overhead and enhance the performance of channel estimation in IRS-enhanced mmWave system, the authors of \cite{estimation3} proposed a channel estimation scheme based on least square estimation with partial on-off and super-resolution network. Although the research on channel estimation of IRS-assisted networks has made rapid progress, channel estimation errors are actually inevitable, and their impact on system performance needs to be considered. Until now, there have been some recent works that studied robust BF designs for IRS-aided networks under imperfect channel state information (CSI). The work \cite{robust1} is the first to study the robust BF based on the imperfect cascaded BS-IRS-user channels at the transmitter. The authors of \cite{robust2} considered the robust active and passive BF co-design in an IRS-aided multiple-input single-output (MISO) communication system, under the general case of correlated CSI errors. Besides, a robust beamforming design based on both bounded CSI error model and statistical CSI error model for PU-related channels in IRS-aided cognitive radio systems was investigated in \cite{robust3}.

Until now, there is a paucity of research on the secure transmission design in the CRNs and mmWave communications, respectively. The authors of \cite{38} focused on enhancing the SR at SU in the IRS-assisted CRNs. Our prior work \cite{39} investigated the SEE maximization problem in the IRS-assisted CRNs to achieve a trade-off between the energy consumption and security. The SR maximization problem in the IRS-aided mmWave communication systems with single IRS and multiple IRSs was investigated in \cite{42} and \cite{43}, respectively. Note that all these existing works \cite{39}, \cite{42}-\cite{43} studied IRS-assisted secure communication networks under the assumption that perfect channel state information (PCSI) of the wiretap link can be obtained. However, in practice, the estimation mismatch makes it very tough to attain PCSI of the wiretap link. Besides, since we can not acquire the position of Eves, and the base station can not interact with Eves, there is almost no way to attain Eves' PCSI. Imperfect CSI will have a devastating impact on the secrecy performance, which shows the importance to design robust secure transmit BF and reflect BF in the IRS-assisted wireless communications. Based on this, the authors of \cite{44} and \cite{45} studied the robust design of IRS-assisted secure wireless systems by adopting a deterministic CSI error model and a statistical CSI error model to characterize the wiretap link CSI uncertainty, respectively. By modeling the imperfect CSI of a multi-antenna Eve as the ellipsoidal bounded uncertainty, \cite{46} investigated the secure transmission in IRS-assisted non-orthogonal multiple access (NOMA) systems. The authors of \cite{47} investigated the SEE maximization problem in IRS-aided MISO networks under imperfect CSI of the wiretap link, modeled by the bounded CSI error. By assuming that the angle of arrival (AoA)-based CSI of wiretap links is imperfect known, a robust secure BF problem was studied in \cite{add5} to make the worst-case SR at its maximum.

However, it should be pointed out that the above works on robust secure design \cite{44}-\cite{47} are only applicable to Rayleigh fading with rich scatters in low frequency band, which are not suitable for the mmWave channels in which the line-of-sight (LoS) component dominates the main component. Although the uncertainty model of the wiretap links in \cite{add5} is suitable for mmWave communications, the robust secure design is not applicable to the IRS-assisted CRNs. In addition, all of the existing works related to IRS-assisted CRNs and IRS-assisted mmWave communication systems are separately investigated. To our best knowledge, this paper is the first to study the robust secure BF design in IRS-assisted mmWave CRNs. In summary, our main contributions are listed as follows:
\begin{itemize}
	\item This is the first research on studying the robust secure transmission problem in the IRS-assisted mmWave CRNs. Due to the estimation mismatch and the passivity of Eves, there is almost no way to attain Eves' PCSI, which motivates our research of the robust secure design.

	\item By using a uniform linear array (ULA) at the CBS and a uniform planar array (UPA) at the IRS, and assuming that angle of departure (AoD)-based imperfect CSIs of wiretap links is known, we formulate a constrained problem to maximize the worst-case achievable secrecy rate (ASR) of the SU subject to the maximum transmit power of CBS, the limited IT of PU and the unit modulus constraint of IRS. The problem is difficult to deal with as a result of the coupling of the transmit BF at the CBS and the reflect BF at the IRS, for which an efficient alternating optimization algorithm is proposed to solve the non-convex problem.
	
	\item The core idea of our proposed robust secure BF design is to exploit the AoD-based Eve channel uncertainty that belongs to a convex hull, which is more accurate in describing the mmWave channel compared with the common deterministic or statistical model \cite{44}-\cite{47}. As for optimizing the transmit BF at the CBS, we propose a heuristic robust transmit BF algorithm to attain the BF vectors analytically. As for the optimization of the reflect BF at the IRS, by means of an auxiliary variable, we transform the non-convex problem into a semi-definite programming (SDP) problem with rank-1 constraint, which is further processed with the aid of an iterative penalty function, and then obtain the optimal reflect BF via CVX.	

	\item The simulation results indicate that compared with other benchmarks, the design of robust secure BF scheme including the transmit BF at the CBS and the reflect BF at the IRS can achieve a satisfying ASR performance, which has a small gap with the optimal solution with PCSI. Moreover, we can find that even if the Eve uncertainty region becomes larger, our proposed robust secure BF scheme can still null the signal leaked to Eves within the channel uncertainty region, validating the effectiveness of our proposed robust secure BF scheme even if there is a relatively big difference between the estimated CSI and actual CSI of wiretap links.
\end{itemize} 

The rest of the paper is organized as follows. Section \uppercase\expandafter{\romannumeral2} introduces the system model followed by the problem formulation. The optimization problem is decoupled into two sub-problems, and is solved in section \uppercase\expandafter{\romannumeral3}. The simulation results are presented in Section \uppercase\expandafter{\romannumeral4}, and this paper is concluded in Section \uppercase\expandafter{\romannumeral5}.

\textit{Notations}: Vectors and matrices are represented by lowercase and uppercase bold typeface letters, respectively. ${\left(\cdot \right)^T}$ and ${\left(\cdot \right)^H}$ indicate the transpose and Hermitian transpose operation, respectively. ${\bf{a}} \otimes {\bf{b}}$ denotes the Kronecker
product of $\bf{a}$ and $\bf{b}$. $\left\| {\cdot} \right\|$ represents the Euclidean norm of a vector. $\left|\cdot\right|$ is the absolute value. $\text{tr}({\bf{X}})$, $\text{rank}({\bf{X}})$, $\left| {\bf{X}} \right|$, $\lambda_{\max}({\bf{X}})$ and $p({\bf{X}})$ denote the trace, rank, determinant,  maximum eigenvalue and corresponding normalized eigenvector of matrix $\bf{X}$, respectively. $\text{diag}({\bf{x}})$ represents the diagonal matrix with ${\bf{x}}$ on its main diagonal. ${\bf{I}}_M$ is the $M\times M$ identity matrix. ${\left[ {\bf{X}} \right]_{i,j}}$ is the $\left( {i,j} \right)$-th element of $\bf{X}$. ${\bf{X}}{\underline \succ} 0$ indicates that $\bf{X}$ is a positive semi-definite matrix. $\left\langle {{\bf{X}},{\bf{Y}}} \right\rangle  = \text{tr}\left( {{{\bf{X}}^H}{\bf{Y}}} \right)$. ${\log _2}\left( \cdot \right)$ and $E\left( {\cdot} \right)$ represent logarithmic function and expectation operator, respectively. $CN\left( {\mu ,{\sigma ^2}} \right)$ stands for the distribution of complex Gaussian random variable with mean $\mu$ and variance $\sigma^2$.

\begin{figure}[t]
	\centering
	\includegraphics[width=9cm]{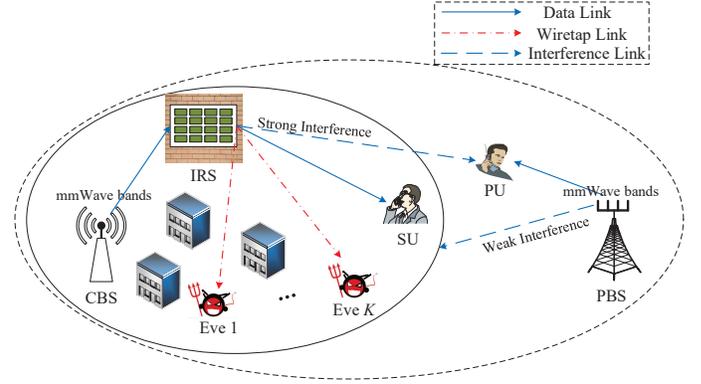}
	\caption{An IRS-assisted cognitive radio network.}
	\label{fig1}
\end{figure}
\section{System Model and Problem Formulation}
As shown in Fig.1, we consider an IRS-assisted mmWave CRN: a CBS exploits spectrum authorized to the PU to communicate with a SU by means of IRS, where \textit{K} Eves attempt to intercept the CBS–SU transmission. Suppose that CBS is equipped with \textit{M} antennas, and SU, PU, and Eves own single antenna. In this paper, the direct links are blocked by obstacles. Thus, an IRS composed of \textit{N} passive reflecting elements is deployed on the facade of a tall building to support blind spot coverage and improve the system performance. It can flexibly adjust the phase of the incident electromagnetic wave. Due to the large path loss, the power of the signals reflected by IRS twice or more is negligible.

All channels in our considered network are supposed to undergo quasi-static flat-fading. It is assumed that the PCSIs of legitimate users are available at the CBS. Nevertheless, Eves usually conceal their location from the CBS, making it hard to attain the PCSIs of wiretap links. Actually, due to the estimation mismatch, the CSI of wiretap links is usually inaccurate \footnote{The AoD-based CSI are available at CBS through cell-ID positioning or satellite GPS and feedback/training sent from the users via backhaul channel, this kind of mechanism has already been suggested and discussed in SUB-S2 \cite{r1}.}. To design a robust secure BF scheme with imperfect CSI of Eves, this paper takes the worst-case Eve channel uncertainty into account.

Next, we will introduce the system model which includes the channel model and signal model, and then formulate the constrained problem.

\subsection{Channel Model}
Suppose that the CBS adopts a $M$-dimensional ULA where the array elements are uniformly placed parallel to the X-axis with inter-element spacing $d$, and the IRS exploits a $N$-dimensional UPA with $N = {N_1} \times {N_2}$, $N_1$ and $N_2$ being the number of array elements uniformly placed along the X-axis and that along the Z-axis with inter-element spacing $d_1$ and $d_2$, respectively. As a result of the quasi-optical and highly directional characteristics of radio wave propagation at mmWave band, a common mmWave channel is modeled as a superposition of a predominant LoS component and a sparse set of single-bounce multipath components \cite{ESI1}. In particular, as illustrated in Fig. 2, we can describe the geometry based mmWave sparse channel between the IRS and user as \cite{ESI1}

\begin{figure}[t]
	\centering
	\includegraphics[width=9cm]{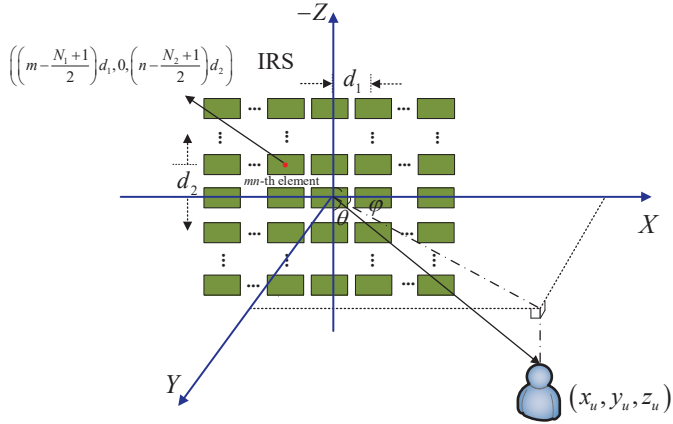}
	\caption{Geometrical relation between IRS and user.}
	\label{fig1}
\end{figure}

\begin{equation}
{{\bf{h}}_{Iv}}{\rm{ = }}\sqrt {\frac{N}{{{L_v}{\rho _v}}}} \sum\limits_{i = 0}^{{L_v} - 1} {{\alpha _{vi}}{{\bf{a}}_{{T_{IRS}}}}\left( {\theta _{vi}^t,\varphi _{vi}^t} \right)} ,v \in \left\{ {{\rm{S}},{\rm{P}},k} \right\},
\end{equation}
where $L_v$ is the number of multipaths between the IRS and user, which includes the LoS $i=0$. $\rho _{v}$ denotes the average path loss between the IRS and user, and $\alpha _{vi}$ is the complex gain. ${{{\bf{a}}_{{T_{IRS}}}}\left( {\theta _{vi}^t,\varphi _{vi}^t} \right)}\in {C^{N \times 1}}$ is the array steering vector at the IRS with $\theta_{vi}^t$ and $\varphi_{vi}^t$ being the vertical and horizontal AoDs of the \emph{i}-th path, respectively.

For convenience, the origin of the IRS array is defined in the center of the array. Denote the location vector of the \emph{mn}-th element (corresponding to the red dot in Fig. 2) and the AoD unit vector as ${\bf{r}}_{mn}^{} = {\left[ {{x_m},0,{z_n}} \right]^T}$ and ${\bf{d}} = {\left[ {\sin \theta \cos \varphi ,\sin \theta \sin \varphi ,\cos \theta } \right]^T}$, respectively. As such, we can calculate the phase delay of the \emph{mn}-th element relative to the origin in the array plane as
\begin{equation}
	\begin{split}
	{\tau _{mn}} =& \frac{{2\pi }}{\lambda } \left\langle {{\bf{r}}_{mn}^T \cdot {\bf{d}}} \right\rangle \\
	=& \frac{{2\pi }}{\lambda } \left( {\left( {m - \left( {{N_1} + 1} \right)/2} \right){d_1}} \right.\sin \theta \cos \varphi \\
	&\left. { + \left( {n - \left( {{N_2} + 1} \right)/2} \right){d_2}\cos \theta } \right),
	\end{split}
\end{equation}
where $\lambda$ is the mmWave wavelength, $m \in \left\{ {1,2, \cdots {N_1}} \right\}$, $n \in \left\{ {1,2, \cdots {N_2}} \right\}$. It should be mentioned that since IRS is usually deployed on the facade of a tall building, IRS can not reflect the signal to the back of IRS. Thus, the range of $\varphi$ should be ${\varphi  \in \left[ {0,\pi } \right]}$. Moreover, the users are generally not higher than the IRS deployed on the facade of high buildings. As a consequence, the range of $\theta$ can be considered as ${\theta  \in \left[ {0,{\pi  \mathord{\left/{\vphantom {\pi  2}} \right.\kern-\nulldelimiterspace} 2}} \right]}$.

Therefore, we can obtain the \emph{mn}-th component of IRS's array steering matrix ${{\bf{A}}_{{T_{IRS}}}}\in {C^{N_1 \times N_2}}$ as
\begin{equation}
	\begin{split}
	&{\left[ {{{\bf{A}}_{{T_{IRS}}}}\left( {\theta ,\varphi } \right)} \right]_{mn}} = \frac{1}{{\sqrt N }}\exp \left( {j{\tau _{mn}}} \right)\\
	=& \frac{1}{{\sqrt N }}\exp \left[ {j\frac{{{\rm{2}}\pi }}{\lambda } \left( {\left( {m - \left( {{N_1} + 1} \right)/2} \right){d_1}\sin \theta \cos \varphi } \right.} \right.\\
	&\left. { + \left. {\left( {n - \left( {{N_2} + 1} \right)/2} \right){d_2}\cos \theta } \right)} \right],
	\end{split}
\end{equation}
which can be further written as
\begin{equation}
{{\bf{A}}_{{T_{IRS}}}}\left( {\theta ,\varphi } \right) = \frac{1}{{\sqrt N }}{{\bf{a}}_{T{h_{IRS}}}}\left( {\theta ,\varphi } \right){\bf{a}}_{T{v_{IRS}}}^H\left( \theta  \right),
\end{equation}
where ${{\bf{a}}_{T{h_{IRS}}}}\left( {\theta ,\varphi } \right)\in {C^{N_1 \times 1}}$ and ${\bf{a}}_{T{v_{IRS}}}\left( \theta  \right)\in {C^{N_2 \times 1}}$ stand for the horizontal and vertical steering vectors of UPA, given by
\begin{equation}
	\begin{split}
	{{\bf{a}}_{T{h_{IRS}}}}\left( {\theta ,\varphi } \right) = \left[ {{e^{ - j\frac{{{\rm{2}}\pi }}{\lambda } \left( {\left( {{N_1} + 1} \right)/2} \right){d_1}\sin \theta \cos \varphi }}, \cdots } \right.\\
	{\left. {,{e^{ + j\frac{{{\rm{2}}\pi }}{\lambda } \left( {\left( {{N_1} + 1} \right)/2} \right){d_1}\sin \theta \cos \varphi }}} \right]^T},
	\end{split}
\end{equation}
\begin{equation}
	\begin{split}
	{\bf{a}}_{T{v_{IRS}}}^{}\left( \theta  \right) = \left[ {{e^{ - j\frac{{{\rm{2}}\pi }}{\lambda } \left( {\left( {{N_2} + 1} \right)/2} \right){d_2}\cos \theta }}, \cdots } \right.\\
	{\left. {,{e^{ + j\frac{{{\rm{2}}\pi }}{\lambda } \left( {\left( {{N_2} + 1} \right)/2} \right){d_2}\cos \theta }}} \right]^T}.
	\end{split}
\end{equation}
To make it simple, the array steering matrix ${{\bf{A}}_{{T_{IRS}}}}$ can be converted into vector form, i.e.,
\begin{equation}
	{{\bf{a}}_{{T_{IRS}}}} = vec\left( {{{\bf{A}}_{{T_{IRS}}}}} \right) =\frac{1}{{\sqrt N }} {{\bf{a}}_{T{h_{IRS}}}}\left( {\theta ,\varphi } \right) \otimes {\bf{a}}_{T{v_{IRS}}}^{}\left( \theta  \right).
\end{equation}

On the other hand, CBS-IRS channel ${\bf{H}}_{CI}$ is given as
\begin{equation}
	\begin{split}
{{\bf{H}}_{CI}}{\rm{ = }}\sqrt {\frac{NM}{{L_I}{\rho_I}}} \sum\limits_{j = 0}^{L_I - 1} {{\alpha _{Ij}}{{\bf{a}}_{{R_{IRS}}}}\left( {\theta _j^r,\varphi _j^r} \right){\bf{a}}_{{T_{CBS}}}^H\left( {\eta _j^t} \right)},
	\end{split}
\end{equation}
where $L_I$ is the number of multipaths between the CBS and IRS, including the LoS $j=0$. $\rho _{I}$ denotes the average path loss between the CBS and IRS, and $\alpha _{Ij}$ is the complex gain.  ${{{\bf{a}}_{{R_{IRS}}}}\left( {\theta _j^r,\varphi _j^r} \right)}\in {C^{N \times 1}}$ is the array steering vector at the IRS with $\theta_j^r$ and $\varphi_j^r$ being the vertical and horizontal AoAs of the \emph{j}-th path, respectively. ${{{\bf{a}}_{{T_{CBS}}}}\left( {\eta _j^t} \right)}\in {C^{M \times 1}}$ represents the array steering vector at the CBS with $\eta_j^t$ being the AoD of the \emph{j}-th path.

Similar to the formula (7), the array steering vector at the IRS can be written as
\begin{equation}
	{{\bf{a}}_{{R_{IRS}}}} =\frac{1}{{\sqrt {N} }} {{\bf{a}}_{R{h_{IRS}}}}\left( {\theta ,\varphi } \right) \otimes {\bf{a}}_{R{v_{IRS}}}^{}\left( \theta  \right),
\end{equation}
where ${{\bf{a}}_{R{h_{IRS}}}}\left( {\theta ,\varphi } \right)\in {C^{N_1 \times 1}}$ and ${\bf{a}}_{R{v_{IRS}}}\left( \theta  \right)\in {C^{N_2 \times 1}}$ represent the horizontal and vertical steering vectors of UPA, respectively. In addition, the array steering vector at the CBS with $M$-dimensional ULA is given by
\begin{equation}
	\begin{split}
{\bf{a}}_{{T_{CBS}}}^{}\left( \eta  \right) =
\frac{1}{{\sqrt M }}\left[ {{e^{ - j\frac{{{\rm{2}}\pi }}{\lambda }\frac{{M + 1}}{2}d\sin \eta }}, \cdots } \right.{\left. {,{e^{ + j\frac{{{\rm{2}}\pi }}{\lambda }\frac{{M + 1}}{2}d\sin \eta }}} \right]^T}.
 \end{split}
\end{equation}

\subsection{Signal Model and Problem Formulation}
As shown in Fig. 1, the CBS transmits signal $s$ following $E\left( {{{\left| s \right|}^2}} \right) = 1$ to SU via IRS. The interference from the primary base station (PBS) to SU and Eves can be regarded as noise due to the long distance between the PBS and secondary network \cite{48}. Then, the signals received at the SU and \textit{k}-th Eve can be respectively written as
\begin{equation}
	{y_S} = {\bf{h}}_{IS}^H{\bf{Q}}{{\bf{H}}_{CI}}{\bf{w}}s + {n_S},
\end{equation}
\begin{equation}
	{y_{k}} = {\bf{h}}_{Ik}^H{\bf{Q}}{{\bf{H}}_{CI}}{\bf{w}}s + {n_{k}},\space k \in \left\{ {1,2,\cdots,K} \right\},
\end{equation}
where $\bf{w}$ represents the transmit BF at the CBS. ${{\bf{H}}_{CI}} \in {{\cal{C}}^{N \times M}}$ denotes the mmWave sparse channel matrix of the CBS-IRS link. $\left\{ {{{\bf{h}}_{IS}},{{\bf{h}}_{Ik}}} \right\}\in {{\cal{C}}^{N \times 1}}$ are the mmWave sparse channel vectors of the IRS to the SU and the \emph{k}-th Eve links, respectively. ${\bf{Q}}{\rm{ }} = {\rm{diag}}\left( {{a _1}{e^{j{\phi _1}}}, \cdots ,{a _n}{e^{j{\phi _n}}}, \cdots ,{a _N}{e^{j{\phi _N}}}} \right)$ stands for the phase shift matrix of the IRS. $\phi _n$ and $a_n$ are the phase shift and the amplitude reflection coefficient of the \textit{n}-th reflecting element, respectively. For simplicity, we set $a_n=1, \forall n$ \cite{45}. Following (11) and (12), the signal-to-noise ratio (SNR) at the SU and $k$-th Eve can be expressed as
\begin{equation}
{\gamma _S} = \frac{{{{\left| {{\bf{q}}_{}^H{{\bf{H}}_S}{\bf{w}}} \right|}^2}}}{{\sigma _S^2}},{\gamma _k} = \frac{{{{\left| {{\bf{q}}_{}^H{{\bf{H}}_k}{\bf{w}}} \right|}^2}}}{{\sigma _k^2}},
\end{equation}
where ${{\bf{H}}_S} = {\rm{diag}}\left( {{\bf{h}}_{IS}^H} \right){{\bf{H}}_{CI}}$, ${{\bf{H}}_k} = {\rm{diag}}\left( {{\bf{h}}_{Ik}^H} \right){{\bf{H}}_{CI}}$, ${\bf{q}} \buildrel \Delta \over = {\left[ {{e^{j{\phi _1}}},{e^{j{\phi _2}}}, \cdots ,{e^{j{\phi _N}}}} \right]^H}$.
In addition, the interference received at the PU from the secondary network can be expressed as
\begin{equation}
\begin{split}
	{y_P} = {\bf{h}}_{IP}^H{\bf{Q}}{{\bf{H}}_{CI}}{\bf{w}}s + {n_P}
	= {\bf{q}}_{}^H{{\bf{H}}_P}{\bf{w}}s + {n_P},
\end{split}
\end{equation}
where ${{\bf{H}}_P} = {\rm{diag}}\left( {{\bf{h}}_{IP}^H} \right){{\bf{H}}_{CI}}$. In (11), (12) and (14), $n_S$,  $n_P$ and $n_{k}$ denote the additive complex white Gaussian noises at the SU, PU, and \emph{k}-th Eve, in which the entries are with zero-mean and variance $\sigma _v^2$, $v \in \left\{ {S,P,k} \right\}$, i.e., ${n_v} \sim CN\left( {0,\sigma _v^2} \right)$. Without loss of generality, the noise power $\sigma_k^2$ is set the same for all $k$, i.e., $\sigma_1^2=\dots=\sigma_K^2=\sigma^2$.

By assuming that \textit{K} Eves cooperatively eavesdrop on the signal sent by the CBS, the ASR of SU can be expressed as
\begin{equation}
	\begin{split}
		{R_{sec }} = { {{R_S} - {R_E}}  } = { {{{\log }_2}\left( {1 + {\gamma _S}} \right) - {{\log }_2}\left( {1 + \sum\limits_{k = 1}^K {{\gamma _{k}}} } \right)}  },
	\end{split}
\end{equation}
where $R_S$ and $R_E$ represent the transmission rate at SU and Eve, respectively. To ensure the secure communication in the IRS-assisted mmWave CRNs, we are meant to make SU's worst-case ASR maximum under the premise that the transmit power of CBS, the IT of PU, and the unit modulus of IRS meet the constraints, which can be mathematically described as 
\begin{align}
	\max\limits_{{\bf{w}},{\bf{q}}}\quad &{R_{sec}}\tag{16a}\\
	{\text{ s.t.}}\quad&\left\| {\bf{w}} \right\|^2 \le P_c^{\max }, \tag{16b}\\
	&{{{{\left| {{\bf{q}}_{}^H{{\bf{H}}_P}{\bf{w}}} \right|}^2}} \mathord{\left/
			{\vphantom {{{{\left| {{\bf{q}}_{}^H{{\bf{H}}_P}{\bf{w}}} \right|}^2}} {\sigma _p^2}}} \right.
			\kern-\nulldelimiterspace} {\sigma _p^2}} \le {\gamma _{th}}, \tag{16c}\\
	&\left| {{{\left[ {\bf{q}} \right]}_n}} \right| = 1,\forall n \in \left\{ {1, \cdots ,N} \right\}=\mathbb N, \tag{16d}
\end{align}
where $P_c^{max}$ and $\gamma_{th}$ represent the maximum transmit power of CBS and predefined interference threshold of PU, respectively. ${{{\left[ {\bf{q}} \right]}_n}}$ stands for the \emph{n}-th element of $\bf{q}$. 

As we can see from (16), the uncertainty CSI of Eve channels makes the problem challenging to deal with. Herein, we adopt an AoD-based uncertainty model to describe the wiretap links. Since we have to investigate the worst-case ASR, we take ${\xi _{vi}} = \frac{{\left| {{\alpha _{vi}}} \right|}}{{\sqrt {{\rho _v}} }}$ as the lower bound ${\xi _{vi,L}}$ in our paper. Then, the channel ${\bf{h}}_{Ik}$ of the \emph{k}-th Eve lies in a given AoD-based range, written by
\setcounter{equation}{16}
\begin{equation}
	\begin{split}
	{\Delta _k} = \left\{ {{{\bf{h}}_{Ik}}\left| {{\theta _k} \in \left[ {{\theta _{k,L}},{\theta _{k,U}}} \right],{\varphi _k} \in \left[ {{\varphi _{k,L}},{\varphi _{k,U}}} \right]} \right.} \right\},
	\end{split}
\end{equation}where the subscripts $k,L$ and $k,U$ represent the lower and upper bounds, respectively. Next, we will consider the uncertainty CSIs of Eves, and propose a worst-case robust secure BF scheme to attain the solution of problem (16).

\section{Robust Secure BF Scheme}
Since the logarithmic function is a monotonically increasing function, we can remove the logarithmic sign of the objective function (16a). By substituting (15) and (17) into (16), we can rewrite problem (16) as a worst-case one, given by
\begin{align}
	\mathop {\max }\limits_{{\bf{w}},{\bf{q}}} \mathop {\min }\limits_{{{\bf{h}}_{Ik}} \in {{{\Delta _k}}}}& \frac{{{{\bf{w}}^H}{\bf{H}}_S^H{\bf{q}}{{\bf{q}}^H}{{\bf{H}}_S}{\bf{w}} + \sigma _S^2}}{{{{\bf{w}}^H}\sum\limits_{k = 1}^K {{\bf{H}}_{k}^H{\bf{q}}{{\bf{q}}^H}{{\bf{H}}_{k}}} {\bf{w}} + \sigma _{}^2}}\tag{18a}\\
	{\text{ s.t.}}\quad&\left\| {\bf{w}} \right\|^2 \le P_c^{\max }, \tag{18b}\\
	&{\left| {{\bf{q}}_{}^H{{\bf{H}}_P}{\bf{w}}} \right|^2} \le I_p^{th}, \tag{18c}\\
	&\left| {{{\left[ {\bf{q}} \right]}_n}} \right| = 1,\forall n \in \left\{ {1, \cdots ,N} \right\}=\mathbb N, \tag{18d}
\end{align}
where $I_p^{th} = {\gamma _{th}}\sigma _P^2$.

Note that problem (18) is non-convex over the coupled $\bf{w}$ and $\bf{q}$, making it challenging to deal with problem (18). Thus, we propose an efficient alternating optimization algorithm to solve $\bf{w}$ and $\bf{q}$ alternatively by fixing the other as constant. To this end, problem (18) is decoupled into two sub-problems, i.e., (19) with given $\bf{q}$ and (32) with given $\bf{w}$.

\subsection{Sub-Problem 1: Optimizing $\bf{w}$ With Given $\bf{q}$}
Denote ${{\bf{F}}_k} = {\rm{diag}}({{\bf{h}}_{I{k}}}){\bf{q}}{{\bf{q}}^H}{\rm{diag}}({\bf{h}}_{I{k}}^H)$, and we can write the sub-problem 1 as
\begin{align}
	\mathop {\max }\limits_{\bf{w}} \mathop {\min }\limits_{{{\bf{F}}_k} \in {\Lambda _k}}& \frac{{{{\bf{w}}^H}{\bf{H}}_S^H{\bf{q}}{{\bf{q}}^H}{{\bf{H}}_S}{\bf{w}} + \sigma _S^2}}{{{{\bf{w}}^H}{\bf{H}}_{CI}^H\sum\limits_{k = 1}^K {{{\bf{F}}_k}{{\bf{H}}_{CI}}} {\bf{w}} + \sigma _{}^2}}\tag{19a}\\
	{\text{ s.t.}}\quad&\left\| {\bf{w}} \right\|^2 \le P_c^{\max }, \tag{19b}\\
	&{\left| {{\bf{q}}_{}^H{{\bf{H}}_P}{\bf{w}}} \right|^2} \le I_p^{th}, \tag{19c}
\end{align}
where $
		{\Lambda _k} = \left\{ {{{\bf{F}}_k} = {\rm{diag}}({{\bf{h}}_{I{k}}}){\bf{q}}{{\bf{q}}^H}{\rm{diag}}({\bf{h}}_{I{k}}^H)} \right.\left| {{\theta _k} \in } \right.\left[ {{\theta _{k,L}},} \right.\\
		\left. {\left. {{\theta _{k,U}}} \right],{\varphi _k} \in \left[ {{\varphi _{k,L}},{\varphi _{k,U}}} \right]} \right\}$.

The transmit BF design at the CBS tends to make the signal reach the IRS to the greatest extent for further reflection. To deal with the intractability of ${\Lambda _k}$, we construct a convex hull of ${\Lambda _k}$ based on weighted sum of $M_k$ discrete samples, which covers all regions of Eve channel as much as possible. As such, we can construct the convex hull of $\Lambda _k$ as \cite{49}
\setcounter{equation}{19}
\begin{equation}
	\begin{split}
{\Psi _k} = \left\{ {\sum\limits_{i = 1}^{{M_k}} {{\mu _{k,i}}{{\bf{F}}_{k,i}}} {\rm{|}}\sum\limits_{i = 1}^{{M_k}} {{\mu _{k,i}} = 1,\;{\mu _{k,i}} \ge 0} } \right\},\;\;\;\forall k,
	\end{split}
\end{equation}
where ${\mu _{k,i}}$ denotes the weight factor. ${{\bf{F}}_{k,i}}$ represents the $i$-th discrete element in $\Lambda _k$. $M_k$ is the number of samples, which should be large enough.

\textit{Proposition 1}: The problem (19) with ${{{\bf{F}}_{k}} \in {\Lambda _k}}$ can be equivalently transformed to that with ${{{\bf{F}}_k} \in {\Psi _k}}$, i.e.,
\begin{equation}
	\begin{split}
\;{\kern 1pt} {\kern 1pt} \mathop {\max }\limits_{\bf{w}} \mathop {\min }\limits_{{{\bf{F}}_k} \in {\Lambda _k}} \frac{{{{\bf{w}}^H}{\bf{H}}_S^H{\bf{q}}{{\bf{q}}^H}{{\bf{H}}_S}{\bf{w}} + \sigma _S^2}}{{{{\bf{w}}^H}{\bf{H}}_{CI}^H\sum\limits_{k = 1}^K {{{\bf{F}}_k}{{\bf{H}}_{CI}}} {\bf{w}} + \sigma _{}^2}} \\= \mathop {\max }\limits_{\bf{w}} \mathop {\min }\limits_{{{\bf{F}}_k} \in {\Psi _k}} \frac{{{{\bf{w}}^H}{\bf{H}}_S^H{\bf{q}}{{\bf{q}}^H}{{\bf{H}}_S}{\bf{w}} + \sigma _S^2}}{{{{\bf{w}}^H}{\bf{H}}_{CI}^H\sum\limits_{k = 1}^K {{{\bf{F}}_k}{{\bf{H}}_{CI}}} {\bf{w}} + \sigma _{}^2}}.
	\end{split}
\end{equation}

\quad\textit{Proof}: Appendix A.

\textit{Proposition 2}: Exchanging the maximum and minimum optimization order on the right side of (21) does not affect the optimization result, namely
\begin{equation}
	\begin{split}
	\mathop {\max }\limits_{\bf{w}} \mathop {\min }\limits_{{{\bf{F}}_k} \in {\Psi _k}} \frac{{{{\bf{w}}^H}{\bf{H}}_S^H{\bf{q}}{{\bf{q}}^H}{{\bf{H}}_S}{\bf{w}} + \sigma _S^2}}{{{{\bf{w}}^H}{\bf{H}}_{CI}^H\sum\limits_{k = 1}^K {{{\bf{F}}_k}{{\bf{H}}_{CI}}} {\bf{w}} + \sigma _{}^2}} \\= \mathop {\min }\limits_{{{\bf{F}}_k} \in {\Psi _k}} \mathop {\max }\limits_{\bf{w}} \frac{{{{\bf{w}}^H}{\bf{H}}_S^H{\bf{q}}{{\bf{q}}^H}{{\bf{H}}_S}{\bf{w}} + \sigma _S^2}}{{{{\bf{w}}^H}{\bf{H}}_{CI}^H\sum\limits_{k = 1}^K {{{\bf{F}}_k}{{\bf{H}}_{CI}}} {\bf{w}} + \sigma _{}^2}}.
	\end{split}
\end{equation}

\quad\textit{Proof}: Appendix B.

According to \textit{Proposition 1} and \textit{Proposition 2}, we can further express the sub-problem 1 as
\begin{align}
	\mathop {\min }\limits_{{{\bf{F}}_k} \in {\Psi _k}} \mathop {\max }\limits_{\bf{w}}& \frac{{{{\bf{w}}^H}{\bf{H}}_S^H{\bf{q}}{{\bf{q}}^H}{{\bf{H}}_S}{\bf{w}} + \sigma _S^2}}{{{{\bf{w}}^H}{\bf{H}}_{CI}^H\sum\limits_{k = 1}^K {{{\bf{F}}_k}{{\bf{H}}_{CI}}} {\bf{w}} + \sigma _{}^2}}\tag{23a}\\
	{\text{ s.t.}}\quad&\left\| {\bf{w}} \right\|^2 \le P_c^{\max }, \tag{23b}\\
	&{\left| {{\bf{q}}_{}^H{{\bf{H}}_P}{\bf{w}}} \right|^2} \le I_p^{th}. \tag{23c}
\end{align}

Noting that $\bf{w}$ is not normalized, we introduce ${\bf{w}} = \sqrt P {\bf{x}}$ which satisfies ${\left\| {\bf{x}} \right\|^2} = 1$. As such, problem (23) can be rewritten as 
\begin{align}
	\mathop {\min }\limits_{{{\bf{F}}_k} \in {\Psi _k}} \mathop {\max }\limits_{{\bf{x}}, P}& \frac{{P{{\bf{x}}^H}{\bf{H}}_S^H{\bf{q}}{{\bf{q}}^H}{{\bf{H}}_S}{\bf{x}} + \sigma _S^2}}{{P{{\bf{x}}^H}{\bf{H}}_{CI}^H\sum\limits_{k = 1}^K {{{\bf{F}}_k}{{\bf{H}}_{CI}}} {\bf{x}} + \sigma _{}^2}}\tag{24a}\\
	{\text{ s.t.}}\quad&{\left\| {\bf{x}} \right\|^2} = 1,P \le P_c^{\max }, \tag{24b}\\
	&\frac{{P{{\bf{x}}^H}{\bf{H}}_P^H{\bf{qq}}_{}^H{{\bf{H}}_P}{\bf{x}}}}{{I_p^{th}}} \le 1. \tag{24c}
\end{align}

By introducing an auxiliary variable $\zeta$ which satisfies $0 \le \zeta  \le 1$ and exploiting the convex hull in (20), we can rewrite the constrained problem (24) as
\begin{align}
	\mathop {\min }\limits_{\left\{ {{\mu _{k,i}}} \right\}} \mathop {\max }\limits_{{\bf{x}},P}& \frac{{P{{\bf{x}}^H}{\bf{H}}_S^H{\bf{q}}{{\bf{q}}^H}{{\bf{H}}_S}{\bf{x}} + \sigma _S^2}}{{P{{\bf{x}}^H}{\bf{H}}_{CI}^H\sum\limits_{k = 1}^K {\sum\limits_{i = 1}^{{M_k}} {{\mu _{k,i}}{{\bf{F}}_{k,i}}} } {\bf{H}}_{CI}^{}{\bf{x}} + \sigma ^2}}\tag{25a}\\
	{\text{ s.t.}}\quad&{\left\| {\bf{x}} \right\|^2} = 1,P \le P_c^{\max }, \tag{25b}\\
	&\frac{{P{{\bf{x}}^H}{\bf{H}}_P^H{\bf{qq}}_{}^H{{\bf{H}}_P}{\bf{x}}}}{{I_p^{th}}} +\zeta= 1. \tag{25c}
\end{align}
Considering that it is still difficult to obtain the analytical transmit BF vector mathematically, we will propose a heuristic approach to attain the analytical solution of sub-problem 1 in the following.

To construct the convex hull $\Psi_k$, we have to generate the key element $\left\{ {{{\bf{F}}_{k,i}}} \right\}$, which can be obtained by selecting $M_k$ representative discrete samples in $\left\{ {\left( {{\theta _k},{\varphi _k}} \right)\left| {\theta_k \in \left[ {{\theta _{k,L}},{\theta _{k,U}}} \right],{\varphi _k} \in \left[ {{\varphi _{k,L}},{\varphi _{k,U}}} \right]} \right.} \right\}$. Hence, for given $P$, $\left\{ {{\mu _{k,i}}} \right\}$ and $\zeta$, we incorporate the constraint (25c) into the denominator of the objective function (25a), yielding a generalized Rayleigh quotient form
\begin{align}
	 \mathop {\max }\limits_{{\bf{x}}}& \frac{{{{\bf{x}}^H}{\bf{Bx}}}}{{{{\bf{x}}^H}{\bf{Ax}}}}\tag{26a}\\
	{\text{ s.t.}}\quad&{\left\| {\bf{x}} \right\|^2} = 1, \tag{26b}
\end{align}
where ${\bf{A}} = P{\bf{H}}_{CI}^H\sum\limits_{k = 1}^K {\sum\limits_{i = 1}^{{M_k}} {{\mu _{k,i}}{{\bf{F}}_{k,i}}} } {\bf{H}}_{CI}^{}+ \frac{{P\sigma _{}^2{\bf{H}}_P^H{\bf{qq}}_{}^H{{\bf{H}}_P}}}{{I_p^{th}}} + \zeta \sigma _{}^2{\bf{I}}_M$, ${\bf{B}} = P{\bf{H}}_S^H{\bf{q}}{{\bf{q}}^H}{{\bf{H}}_S} + \sigma _S^2{\bf{I}}_M$. The optimization variable and optimization objective function of problem (26) can be solved by making use of the generalized Rayleigh quotient theorem, respectively given by \setcounter{equation}{26}
\begin{equation}
	\begin{split}
		{\bf{x}}\left( {P,{\mu _{k,i}},\zeta } \right) = p\left( {{{\bf{A}}^{ - 1}}{\bf{B}}} \right),\\
\gamma \left( {P,{\mu _{k,i}},\zeta } \right) = {\lambda _{\max }}\left( {{{\bf{A}}^{ - 1}}{\bf{B}}} \right),
	\end{split}
\end{equation}

Since the solution obtained in (27) is solved with given $P$, $\left\{ {{\mu _{k,i}}} \right\}$ and $\zeta$, we have to calculate them to make the optimal objective function achieved. Considering that SU's received gain after BF is definitely greater than Eves', i.e., ${{\bf{x}}^H}{\bf{H}}_S^H{\bf{q}}{{\bf{q}}^H}{{\bf{H}}_S}{\bf{x}} > {{\bf{x}}^H}{\bf{H}}_{CI}^H\sum\limits_{k = 1}^K {\sum\limits_{i = 1}^{{M_k}} {{\mu _{k,i}}{{\bf{F}}_{k,i}}} } {\bf{H}}_{CI}^{}{\bf{x}}$, the objective function (26a) increases monotonically with respect to $P$, urging us to take the maximum value of $P$. Note that $P$ is subject to the constraints (24b) and (24c), and we can attain
\begin{equation}
	\begin{split}
		P = \min \left( {{P_{\max }},\frac{{I_p^{th}}}{{{{\bf{x}}^H}{\bf{H}}_P^H{\bf{qq}}_{}^H{{\bf{H}}_P}{\bf{x}}}}} \right).
	\end{split}
\end{equation}
Then, by substituting (28) into (25c), we can obtain
\begin{equation}
	\begin{split}
\zeta  = 1 - \frac{{P{{\bf{x}}^H}{\bf{H}}_P^H{\bf{qq}}_{}^H{{\bf{H}}_P}{\bf{x}}}}{{I_p^{th}}}.
	\end{split}
\end{equation}
After $P$ and $\zeta$ have been calculated to make $\frac{{P{{\bf{x}}^H}{\bf{H}}_S^H{\bf{q}}{{\bf{q}}^H}{{\bf{H}}_S}{\bf{x}} + \sigma _S^2}}{{P{{\bf{x}}^H}{\bf{H}}_{CI}^H\sum\limits_{k = 1}^K {\sum\limits_{i = 1}^{{M_k}} {{\mu _{k,i}}{{\bf{F}}_{k,i}}} } {\bf{H}}_{CI}^{}{\bf{x}} + \sigma ^2}}$ at its maximum with given $\mu_{k,i}$, $\mu_{k,i}$ also need to be selected to make the minimum objective function achieved with calculated $P$ and $\zeta$. With the help of Cauchy-Schwarz inequality, we can attain
\begin{equation}
	\begin{split}
&{\left( {\sum\limits_{i = 1}^{{M_k}} {{\mu _{k,i}}{{\bf{x}}^H}{\bf{H}}_{CI}^H{{\bf{F}}_{k,i}}} {\bf{H}}_{CI}^{}{\bf{x}}} \right)^2} \\
\le& \left( {\sum\limits_{i = 1}^{{M_k}} {\mu _{k,i}^2} } \right)\left( {\sum\limits_{i = 1}^{{M_k}} {{{\left( {{{\bf{x}}^H}{\bf{H}}_{CI}^H{{\bf{F}}_{k,i}}{\bf{H}}_{CI}^{}{\bf{x}}} \right)}^2}} } \right),
	\end{split}
\end{equation}
if and only if $\frac{{{\mu _{k,1}}}}{{{{\bf{x}}^H}{\bf{H}}_{CI}^H{{\bf{F}}_{k,1}}{\bf{H}}_{CI}^{}{\bf{x}}}} = \frac{{{\mu _{k,2}}}}{{{{\bf{x}}^H}{\bf{H}}_{CI}^H{{\bf{F}}_{k,2}}{\bf{H}}_{CI}^{}{\bf{x}}}} =  \cdots {\kern 1pt} {\kern 1pt}  = \frac{{{\mu _{k,{M_k}}}}}{{{{\bf{x}}^H}{\bf{H}}_{CI}^H{{\bf{F}}_{k,{M_k}}}{\bf{H}}_{CI}^{}{\bf{x}}}}$, the equation holds. To investigate the worst-case ASR in (25a), we have to let ${\sum\limits_{i = 1}^{{M_k}} {{\mu _{k,i}}{{\bf{x}}^H}{\bf{H}}_{CI}^H{{\bf{F}}_{k,i}}} {\bf{H}}_{CI}^{}{\bf{x}}}$ take the maximum value, i.e., make the formula (30) take the equal sign. Denote $\frac{{{\mu _{k,1}}}}{{{{\bf{x}}^H}{\bf{H}}_{CI}^H{{\bf{F}}_{k,1}}{\bf{H}}_{CI}^{}{\bf{x}}}} = \frac{{{\mu _{k,2}}}}{{{{\bf{x}}^H}{\bf{H}}_{CI}^H{{\bf{F}}_{k,2}}{\bf{H}}_{CI}^{}{\bf{x}}}} =  \cdots {\kern 1pt} {\kern 1pt}  = \frac{{{\mu _{k,{M_k}}}}}{{{{\bf{x}}^H}{\bf{H}}_{CI}^H{{\bf{F}}_{k,{M_k}}}{\bf{H}}_{CI}^{}{\bf{x}}}}=z$, and we can write ${\mu _{k,i}} = z{{\bf{x}}^H}{\bf{H}}_{CI}^H{{\bf{F}}_{k,i}}{\bf{H}}_{CI}^{}{\bf{x}}$, which is further substituted into the constraint $\sum\limits_{i = 1}^{{M_k}} {{\mu _{k,i}} = 1} $, yielding $z = \frac{1}{{\sum\limits_{i = 1}^{{M_k}} {{{\bf{x}}^H}{\bf{H}}_{CI}^H{{\bf{F}}_{k,i}}} {\bf{H}}_{CI}^{}{\bf{x}}}}$. Thus, ${{\mu _{k,i}}}$ should be selected as
\begin{equation}
	\begin{split}
{\mu _{k,i}} = \frac{{{{\bf{x}}^H}{\bf{H}}_{CI}^H{{\bf{F}}_{k,i}}{\bf{H}}_{CI}^{}{\bf{x}}}}{{\sum\limits_{i = 1}^{{M_k}} {{{\bf{x}}^H}{\bf{H}}_{CI}^H{{\bf{F}}_{k,i}}} {\bf{H}}_{CI}^{}{\bf{x}}}}.
	\end{split}
\end{equation}

Finally, we summarize our heuristic algorithm as Algorithm 1, which can attain the robust transmit BF at the CBS analytically.
	\linespread{1}
\begin{algorithm}[t]
	{\footnotesize{
	\caption{The Proposed Heuristic Robust Transmit BF Algorithm for Solving Sub-Problem 1}
	\LinesNumbered 
	\KwIn{${\bf{q}}$, ${\bf{H}}_{CI}$, ${\bf{h}}_S$, ${\bf{h}}_P$, ${{I}}_{p}^{th}$, $P_c^{max}$}
	\KwOut{Robust transmit BF ${\bf{w}}^*$ at the CBS}
		Initialize $n=0$, $P^0=P_{\max}$, $\zeta^0=0$, $\mu _{k,i}^0 = \frac{1}{{{M_k}}}$\;
		Construct the convex hull $\Psi_{k}$\;
		Calculate ${\bf{x}}_0$ and $\gamma^0$ by making use of (27) with $P^0$, $\zeta^0$ and $\mu _{k,i}^0$\;
		\ShowLn	
		\Repeat{$\left| {{\gamma^{n}} - {\gamma^{n-1}}} \right| \le \varepsilon $}{
			(\rmnum{1}) $n := n+1$\;
			(\rmnum{2}) Calculate $P^n = \min \left( {{P_c^{max}},\frac{{I_p^{th}}}{{{\bf{x}}_{n - 1}^H{\bf{H}}_P^H{\bf{qq}}_{}^H{{\bf{H}}_P}{{\bf{x}}_{n - 1}}}}} \right)$\;
			(\rmnum{3}) Update ${\zeta ^n}: = 1 - \frac{{P^n{\bf{x}}_{n - 1}^H{\bf{H}}_P^H{\bf{qq}}_{}^H{{\bf{H}}_P}{{\bf{x}}_{n - 1}}}}{{I_p^{th}}}$\;
			(\rmnum{4}) Compute $\mu _{k,i}^n: = \frac{{{\bf{x}}_{n - 1}^H{\bf{H}}_{CI}^H{{\bf{F}}_{k,i}}{\bf{H}}_{CI}^{}{{\bf{x}}_{n - 1}}}}{{\sum\limits_{i = 1}^{{M_k}} {{\bf{x}}_{n - 1}^H{\bf{H}}_{CI}^H{{\bf{F}}_{k,i}}} {\bf{H}}_{CI}^{}{{\bf{x}}_{n - 1}}}}$\;
			(\rmnum{5}) Calculate ${\bf{x}}_n$ and $\gamma^n$ with the help of (27)\;			

		}	
				
		\ShowLn
		Exploit (27) to obtain ${\bf{x}}^*$ with $P^n$, $\zeta^n$ and $\mu_{k,i}^n$\;
		Compute ${\bf{w}}^*=\sqrt {{P^n}} {\bf{x}}^*$\;

	\SetKwProg{Fn}{Function}{}{end}}}

\end{algorithm}

\subsection{Sub-Problem 2: Optimizing $\bf{q}$ With Given $\bf{w}$}
By means of a new variable ${\bf{\Theta}}={\bf{q}}{\bf{q}}^H$, problem (18) thereby can be equivalently expressed as
\begin{align}
	\mathop {\max }\limits_{\bf{\Theta }} \mathop {\min }\limits_{{{\bf{h}}_{Ik}} \in {\Delta _k}}& \frac{{\text{tr}\left( {{{\bf{H}}_S}{\bf{w}}{{\bf{w}}^H}{\bf{H}}_S^H{\bf{\Theta }}} \right) + \sigma _S^2}}{{\sum\limits_{k = 1}^K {\text{tr}\left( {{{\bf{H}}_{k}}{\bf{w}}{{\bf{w}}^H}{\bf{H}}_{k}^H{\bf{\Theta }}} \right)}  + \sigma _{}^2}}\tag{32a}\\
	{\text{ s.t.}}\quad
	&\text{tr}\left( {{{\bf{H}}_P}{\bf{w}}{{\bf{w}}^H}{\bf{H}}_P^H{\bf{\Theta }}} \right) \le I_p^{th}, \tag{32b}\\
	&{\left[ {\bf{\Theta }} \right]_{n,n}} = 1, \forall n \in \left\{ {1, \cdots ,N} \right\}=\mathbb N, \tag{32c}\\
	&\text{rank}\left( {\bf{\Theta }} \right) = 1\tag{32d}.
\end{align}
Denote ${{\bf{G}}_k} = {\rm{diag}}({\bf{h}}_{Ik}^H){{\bf{H}}_{CI}}{\bf{w}}{{\bf{w}}^H}{\bf{H}}_{CI}^H{\rm{diag}}({{\bf{h}}_{Ik}})$, and we can write the sub-problem 2 as
\begin{align}
\mathop {\max }\limits_{\bf{\Theta }} \mathop {\min }\limits_{{{\bf{G}}_k} \in {\Xi _k}} &\frac{{\text{tr}\left( {{{\bf{H}}_S}{\bf{w}}{{\bf{w}}^H}{\bf{H}}_S^H{\bf{\Theta }}} \right) + \sigma _S^2}}{{\sum\limits_{k = 1}^K {\text{tr}\left( {{{\bf{G}}_k}{\bf{\Theta }}} \right)}  + \sigma _{}^2}}\tag{33a}\\
	{\text{ s.t.}}\quad
	&(32b)-(32d), \tag{33b}
\end{align}
where $
	{\Xi _k} = \left\{ {{{\bf{G}}_k} = {\rm{diag}}({\bf{h}}_{Ik}^H){{\bf{H}}_{CI}}{\rm{w}}{{\rm{w}}^H}{\bf{H}}_{CI}^H{\rm{diag}}({{\bf{h}}_{Ik}})} \right. \\ \left| {{\theta _k} \in \left[ {{\theta _{k,L}},{\theta _{k,U}}} \right],} \right.\left. {{\varphi _k} \in \left[ {{\varphi _{k,L}},{\varphi _{k,U}}} \right]} \right\}$.

The design of reflect BF at the IRS aims to make signal reflected to possible uncertainty region of Eve channel as small as possible, so as to maximum SU's worst-case ASR. As such, we construct the convex hull of $\Xi _k$ as \cite{49}
\setcounter{equation}{33}
\begin{equation}
	\begin{split}
		{\Upsilon  _k} = \left\{ {\sum\limits_{i = 1}^{{M_k}} {{\chi _{k,i}}{{\bf{G}}_{k,i}}} {\rm{|}}\sum\limits_{i = 1}^{{M_k}} {{\chi _{k,i}} = 1,\;{\chi  _{k,i}} \ge 0} } \right\},\;\;\;\forall k,
	\end{split}
\end{equation}
where ${\chi _{k,i}}$ denotes the weight factor. ${{\bf{G}}_{k,i}}$ represents the $i$-th discrete elements in $\Xi _k$. $M_k$ is the number of samples.

By employing the \emph{Proposition 1}, problem (33) can be rewritten as 
\begin{align}
\mathop {\max }\limits_{\bf{\Theta }} \mathop {\min }\limits_{{{\bf{G}}_k} \in {\Upsilon _k}} &\frac{{\text{tr}\left( {{{\bf{H}}_S}{\bf{w}}{{\bf{w}}^H}{\bf{H}}_S^H{\bf{\Theta }}} \right) + \sigma _S^2}}{{\sum\limits_{k = 1}^K {\text{tr}\left( {{{\bf{G}}_k}{\bf{\Theta }}} \right)}  + \sigma _{}^2}}\tag{35a}\\
	{\text{ s.t.}}\quad
	&(32b)-(32d)\tag{35b}.
\end{align}

For the convenience of the next transformation, we first take the reciprocal of (35a), given by
\begin{align}
\mathop {\min }\limits_{\bf{\Theta }} \mathop {\max }\limits_{{{\bf{G}}_k} \in {\Upsilon _k}} &{\rm{ }}\frac{{\sum\limits_{k = 1}^K {\text{tr}\left( {{{\bf{G}}_k}{\bf{\Theta }}} \right)}  + \sigma _{}^2}}{{\text{tr}\left( {{{\bf{H}}_S}{\bf{w}}{{\bf{w}}^H}{\bf{H}}_S^H{\bf{\Theta }}} \right) + \sigma _S^2}}\tag{36a}\\
	{\text{ s.t.}}\quad
	&(32b)-(32d)\tag{36b}.
\end{align}

Then, by means of two introduced variables $t$ and $r$, a positive semi-definite matrix $\bf{R}$ which satisfies ${\bf{R}}=r{\bf{\Theta}}$, as well as the Charnes-Cooper transformation \cite{50}, we can transform (36) into a SDP problem, given by
\begin{align}
		\mathop {\min }\limits_{{\bf{R}} {\underline \succ}0,r \ge 0}  &t\tag{37a}\\
	{{\rm{ s}}{\rm{.t}}{\rm{.}}\quad }&\mathop {\max }\limits_{{{\bf{G}}_k} \in {\Upsilon _k}} \sum\limits_{k = 1}^K {\text{tr}\left( {{{\bf{G}}_k}{\bf{R}}} \right)}  + r\sigma _{}^2 \le t,\tag{37b}\\
	&\text{tr}\left( {{{\bf{H}}_S}{\bf{w}}{{\bf{w}}^H}{\bf{H}}_S^H{\bf{R }}} \right) + r\sigma _S^2 \ge 1,\tag{37c}\\
	&\text{tr}\left( {{{\bf{H}}_P}{\bf{w}}{{\bf{w}}^H}{\bf{H}}_P^H{\bf{R}}} \right) \le rI_p^{th} ,\tag{37d}\\
	&{\left[ {\bf{R}} \right]_{n,n}} = r,\forall n \in \left\{ {1, \cdots ,N} \right\} = \mathbb N,\tag{37e}\\
	&\text{rank}({\bf{R}}) = 1.\tag{37f}
\end{align}

\textit{Proposition 3}: The optimization problem (37) is completely equivalent to the problem (36), indicating that the solutions of problems (36) and (37) are the same, given by ${{\bf{\Theta }}^*} = {{{{\bf{R}}^*}} \mathord{\left/{\vphantom {{{{\bf{R}}^*}} {{r^*}}}} \right.\kern-\nulldelimiterspace} {{r^*}}}$.

\quad\textit{Proof}: Appendix C.

Construct the convex hull ${\Upsilon _k}$ in a similar way to Section \textit{A}. Then, with the help of the convex hull (34), we can rewrite the constraint (37b) as
\setcounter{equation}{37}
\begin{equation}		
	\begin{split}
\mathop {\max }\limits_{\left\{ {{\chi _{k,i}}} \right\}} \sum\limits_{k = 1}^K {\text{tr}\left( {\sum\limits_{i = 1}^{{M_k}} {{\chi _{k,i}}{{\bf{G}}_{k,i}}{\bf{R}}} } \right)}  + r\sigma _{}^2 \le t,
	\end{split}
\end{equation}
In what follows, we first discuss to solve the problem (37) with given $\left\{ {{\chi _{k,i}}} \right\}$. Then, we calculate $\left\{ {{\chi _{k,i}}} \right\}$ to make the maximum value on the left side of the constraint (38) achieved, i.e., to obtain the maximum ASR in the worst case.

It is obvious that the constraint (37f) makes the optimization problem hard to solve. The traditional semi-definite relaxation (SDR) method usually ignores (37f) to make the problem simplified, and the optimal solution selected among randomly generated rank-1 feasible solutions can be regarded  as an approximately best solution \cite{51}. However, there is probably no optimal solution to the initial SDP problem among the feasible solutions in random space. Even if there is, the chosen rank-1 solution is very likely to be a sub-optimal solution. What's worse, the obtained solution may deviate greatly with the optimal solution. 

\textit{Proposition 4}: ${\rm{rank}}\left( {\bf{R}} \right) = 1 \Leftrightarrow \text{tr}\left( {\bf{R}} \right) - {\lambda _{\max }}\left( {\bf{R}} \right) \le 0$ for any matrix ${\bf{R}}{\underline \succ}0$.

\quad\textit{Proof}: Appendix D.

According to the {\textit{Proposition 4}}, the constrained problem (37) is further reformulated as
\begin{align}
	\mathop {\min }\limits_{{\bf{R}} {\underline \succ}0,r \ge 0}  &t\tag{39a}\\
	{{\rm{ s}}{\rm{.t}}{\rm{.}}\quad }&(37c)-(37e), (38), \tag{39b}\\
	&\text{tr}\left( {\bf{R}} \right) - {\lambda _{\max }}\left( {\bf{R}} \right)\le0.\tag{39c}
\end{align}

Note that the inequality $\text{tr}\left( {\bf{R}} \right) \ge {\lambda _{\max }}\left( {\bf{R}} \right) $ always holds for any matrix ${\bf{R}}{\underline \succ}0$. Thus, our purpose is to make $\text{tr}\left( {\bf{R}} \right) - {\lambda _{\max }}\left( {\bf{R}} \right)$ as small as possible (approaching zero). With the help of penalty item method, the constraint (39c) can be incorporated into the objective function (39a), yielding
\begin{align}
	\min\limits_{{\bf{R}}{\underline \succ}0, r\ge0, t}\quad &{t }+\rho(\text{tr}\left( {\bf{R}} \right) - {\lambda _{\max }}\left( {\bf{R}} \right))\tag{40a}\\
	{\text{ s.t.}}\quad
	&(37c)-(37e), (38), \tag{40b}
\end{align}where the penalty coefficient $\rho$ should be large enough to make $\text{tr}\left( {\bf{R}} \right) - {\lambda _{\max }}\left( {\bf{R}} \right)$ as small as possible. As we can see, the objective function (40a) is concave, making the problem (40) a concave function minimization problem over a convex set, i.e., a concave programming. Moreover, considering that ${\lambda _{\max }}\left( {\bf{R}} \right)$ is a non-smooth function, we can adopt the sub-gradient of the non-smooth function, which is defined as $\partial  {\lambda _{\max }}\left( {\bf{X}} \right) = {{\bf{x}}_{\max }}{\bf{x}}_{\max }^H$. Then, we have \cite{52}
\setcounter{equation}{40}
\begin{equation}		
	\begin{split}
		{\lambda _{\max }}\left( {\bf{X}} \right) - {\lambda _{\max }}\left( {\bf{R}} \right) \ge \left\langle {{{\bf{r}}_{\max }}{\bf{r}}_{\max }^H,{\bf{X}} - {\bf{R}}} \right\rangle ,\forall {\bf{X}} {\underline \succ} 0.
	\end{split}
\end{equation}

Next, by employing the maximum eigenvalue as well as the corresponding unit eigenvector ${\bf{r}}^{(m)}$ to initialize the feasible solution ${\bf{R}}^{(m)}$, which satisfies the constraint (40b), a SDP problem can be written as 
\begin{align}
	\min\limits_{{\bf{R}}{\underline \succ}0, r\ge0, t}\quad &{t + \rho \left[ {\text{tr}\left( {\bf{R}} \right) - \left\langle {{{\bf{r}}^{(m)}}{\bf{r}}_{}^{(m)H},{\bf{R}}} \right\rangle } \right]}\tag{42a}\\
	{\text{ s.t.}}\quad
	&(37c)-(37e), (38),\tag{42b}
\end{align}
The problem (42) can provide the optimal solution for ${\bf{R}}^{(m+1)}$, which produces a smaller objective value (42a) than that produced by ${\bf{R}}^{(m)}$. In specific, we suppose that ${\bf{R}}^{(m+1)}$ is the optimal solved solution of (42), yielding
\setcounter{equation}{42}
\begin{equation}
	\begin{split}
		f\left( {{{\bf{R}}^{\left( {m + 1} \right)}}} \right) &= t + \rho \left[ {\text{tr}\left( {{{\bf{R}}^{\left( {m + 1} \right)}}} \right) - {\lambda _{\max }}\left( {{{\bf{R}}^{\left( {m + 1} \right)}}} \right)} \right]\\
		&\le t + \rho \left[ {\text{tr}\left( {{{\bf{R}}^{\left( {m + 1} \right)}}} \right) - {\lambda _{\max }}\left( {{{\bf{R}}^{\left( m \right)}}} \right)} \right.\\
		&\quad\left. { - \left\langle {{{\bf{r}}^{(m)}}{\bf{r}}_{}^{(m)H},{{\bf{R}}^{\left( {m + 1} \right)}} - {{\bf{R}}^{\left( m \right)}}} \right\rangle } \right]\\
		&= t + \rho \left[ {\text{tr}\left( {{{\bf{R}}^{\left( {m + 1} \right)}}} \right) - \left\langle {{{\bf{r}}^{(m)}}{\bf{r}}_{}^{(m)H},{{\bf{R}}^{\left( {m + 1} \right)}}} \right\rangle } \right.\\
		&\quad\left. { + \left\langle {{{\bf{r}}^{(m)}}{\bf{r}}_{}^{(m)H},{{\bf{R}}^{\left( m \right)}}} \right\rangle  - {\lambda _{\max }}\left( {{{\bf{R}}^{\left( m \right)}}} \right)} \right]\\
		&\le t + \rho \left[ {\text{tr}\left( {{{\bf{R}}^{\left( m \right)}}} \right) - \left\langle {{{\bf{r}}^{(m)}}{\bf{r}}_{}^{(m)H},{{\bf{R}}^{\left( m \right)}}} \right\rangle } \right.\\
		&\quad\left. { + \left\langle {{{\bf{r}}^{(m)}}{\bf{r}}_{}^{(m)H},{{\bf{R}}^{\left( m \right)}}} \right\rangle  - {\lambda _{\max }}\left( {{{\bf{R}}^{\left( m \right)}}} \right)} \right]\\
		&= t + \rho \left[ {\text{tr}\left( {{{\bf{R}}^{\left( m \right)}}} \right) - {\lambda _{\max }}\left( {{{\bf{R}}^{\left( m \right)}}} \right)} \right]\\
		&= f\left( {{{\bf{R}}^{\left( m \right)}}} \right),
	\end{split}
\end{equation}
which verifies the iterative procedure.
	\linespread{1}
\begin{algorithm}[t]
	{\footnotesize{
	\caption{The Proposed Iterative Robust Reflect BF Algorithm for Solving Sub-Problem 2}
	\LinesNumbered 
	\KwIn{${\bf{w}}$, ${{\bf{H}}_{CI}}$, ${\bf{h}}_S$, ${\bf{h}}_P$, $I_p^{th}$}
	\KwOut{Robust reflect BF ${\bf{q}}^*$ at the IRS, and the maximum ASR $R^*$ in the worst case}
	{{Initialize}} $j=0$, $t^0=0$, $\rho=10$, $\chi _{k,i}^0 = \frac{1}{{{M_k}}}$\; 
	Construct the convex hull $\Upsilon _k$\;
	Compute ${\bf{R}}^{(0)}$ satisfying (37c)-(37e) and (38)\;
	\Repeat{$\left| {{t^{(j)}} - {t^{(j-1)}}} \right| \le \varepsilon $}{
		Set $m=0$\;
		\While{$\left| {{\rm{tr}}\left( {{{\bf{R}}^{\left( m \right)}}} \right) - {\lambda _{\max }}\left( {{{\bf{R}}^{\left( m \right)}}} \right)} \right| > \varepsilon $}{
			Find the optimal solution ${\bf{R}}^{(m+1)}$, ${{r}}^{(m+1)}$ and ${{t}}^{(m+1)}$ of problem (42) by using CVX\;
			\eIf{${{\bf{R}}^{\left( {m + 1} \right)}} \approx {{\bf{R}}^{\left( m \right)}}$}{
				Set $\rho:=2\rho$\;
			}{
				Set $m:=m+1$\;
				Set $\rho=10$\;
			}
		}
		Set $j:=j+1$\;
		Set ${{\bf{R}}^{(j)}}: = {{\bf{R}}^{(m)}}$, $r^{(j)}:=r^{(m)}$, $t^{(j)}:=t^{(m)}$\;
		Compute ${\chi _{k,i}}: = \frac{{{\rm{tr}}\left( {{{\bf{G}}_{k,i}}{{\bf{R}}^{\left( j \right)}}} \right)}}{{\sum\limits_{i = 1}^{{M_k}} {{\rm{tr}}\left( {{{\bf{G}}_{k,i}}{{\bf{R}}^{\left( j \right)}}} \right)} }}$\;				
	}	
	
	Calculate ${{\bf{\Theta }}^*} = {{{{\bf{R}}^{\left( j \right)}}} \mathord{\left/
			{\vphantom {{{{\bf{R}}^{\left( m \right)}}} {{r^{\left( m \right)}}}}} \right.
			\kern-\nulldelimiterspace} {{r^{\left( j \right)}}}}$\;
	Obtain the optimal robust reflect BF ${\bf{q}}^*$ through Cholesky decomposition over ${{\bf{\Theta }}^*}$, and the maximum ASR $R^*={\log _2}\left( {\frac{{\sigma ^2}}{{\sigma _{S}^2}}{t^{\left( j \right)}}} \right)$.}}
\end{algorithm}

Until now, we have solved the sub-problem 2 through (42) with given $\left\{ {{\chi _{k,i}}} \right\}$. Next, we need to calculate $\left\{ {{\chi _{k,i}}} \right\}$ to make the maximum value on the left side of the constraint (38) achieved, i.e., to obtain the maximum ASR in the worst case. Rewrite the constraint (38) as 
\begin{equation}
	\begin{split}
\mathop {\max }\limits_{\left\{ {{\chi _{k,i}}} \right\}} \sum\limits_{k = 1}^K {\sum\limits_{i = 1}^{{M_k}} {{\chi _{k,i}}{\rm{tr}}\left( {{{\bf{G}}_{k,i}}{\bf{R}}} \right)} }  + r\sigma^2 \le t.
	\end{split}
\end{equation}
According to Cauchy-Schwarz inequality, we can calculate $\chi_{k,i}$ to achieve the maximum value of ${\sum\limits_{i = 1}^{{M_k}} {{\chi _{k,i}}{\rm{tr}}\left( {{{\bf{G}}_{k,i}}{\bf{R}}} \right)} }$ in a similar manner as (30)-(31), given by
\begin{equation}
	\begin{split}
{\chi _{k,i}} = \frac{{{\rm{tr}}\left( {{{\bf{G}}_{k,i}}{\bf{R}}} \right)}}{{\sum\limits_{i = 1}^{{M_k}} {{\rm{tr}}\left( {{{\bf{G}}_{k,i}}{\bf{R}}} \right)} }}
	\end{split}
\end{equation}

As a result, based on the above analysis, we can divide the solution process into two steps. Firstly, solve the problem (42) to obtain $\bf{R}$ through CVX with given $\left\{ {{\chi _{k,i}}} \right\}$, and then update $\left\{ {{\chi _{k,i}}} \right\}$ according to the solved $\bf{R}$. The selection of the penalty coefficient $\rho$ is important for the computational efficiency. Algorithm 2 shows the detailed solution process of sub-problem 2, which includes the choice of the penalty coefficient $\rho$.

\subsection{The Overall Algorithm}
With the help of generalized Rayleigh entropy, Charnes-Cooper transformation, as well as penalty function, we propose an efficient alternating optimization algorithm to attain the maximum ASR in the worst case and corresponding optimal robust BF vectors $({\bf{w}^*}, {\bf{q}^*})$. Algorithm 3 illustrates the flow of the overall algorithm.
	\linespread{1}
\begin{algorithm} [t]
	{\footnotesize{
	\caption{ Alternating Optimization Algorithm for Solving Problem (18)}		
	\LinesNumbered 
	\KwIn{${{\bf{H}}_{CI}}$, ${\bf{h}}_S$, ${\bf{h}}_P$, $P_c^{max}$, $I_p^{th}$}
	\KwOut{Optimal transmit BF ${\bf{w}}^*$ at the CBS, optimal reflect BF ${\bf{q}}^*$ at the IRS, and the maximum ASR $R^*$ in the worst case}
	Initialize $p=0$, $R^{(0)}=0$, ${\bf{q}}^*$\;
	\ShowLn
	\Repeat{$\left| {{R^{(p)}} - {R^{(p-1)}}} \right| \le \varepsilon$} 
	{
		Set ${\bf{q}}$:=${\bf{q}}^*$\;
		Perform {\bf{Algorithm 1}} with given ${\bf{q}}$ to obtain ${\bf{w}}^*$\;
		Set ${\bf{w}}$:=${\bf{w}}^*$\;
		Perform {\bf{Algorithm 2}} with given ${\bf{w}}$ to obtain ${\bf{q}}^*$ and $R^*$\;
		Update ${R^{(p + 1)}} := R^*$\;
		Set $p:=p+1$\;	
	}
	Obtain \\
	(i) the robust optimal transmit BF ${{\bf{w}^*}}$\;
	(ii) the robust optimal reflect BF ${\bf{q}}^{*}$\;
	(iii) the maximum ASR $R^*=R^{(p)}$ in the worst case.}}
\end{algorithm}

\subsection{Complexity Analysis}
The main complexity of sub-problem 1 comes from matrix operation \cite{Co1}. As such, the complexity of sub-problem 1 is $O\left\{ {16{N^2}M + 16{M^2}N + } \right.{I_1}\left[ {K{M_k}{N^2} + 8{N^2}M + 8{M^2}N + } \right.$ $\left. {\left. {142{M^3}} \right]} \right\}$, where $I_1$ denotes the iteration number of sub-problem 1. According to \cite{Co2}, the computational complexity per-iteration in sub-problem2 is mainly caused by the number of optimization variables, the number of linear matrix inequality constraints. Note that problem (42) has $N^2$ design variables and 2 slack variables, $N+3$ LMI constraints of size 1. As a result, the computational complexity of sub-problem 2 is $O\left\{ {{I_2}\left[ {\sqrt {N + 3} } \right.} \right.\left. {\left. {n\left( {\left( {N + 3} \right)\left( {n + 1} \right) + {n^2}} \right)} \right]} \right\}$, where $n = O\left( {\sqrt {{N^2} + 2} } \right)$ and $I_2$ is the iteration number of sub-problem 2. Consequently, the overall complexity of proposed algorithm is $O\left\{ {{I_{outer}}\left[ {16{N^2}M + 16{M^2}N} \right.} \right. + {I_1}\left( {K{M_k}{N^2}} \right.$ $\left. { + 8{N^2}M + 8{M^2}N + 142{M^3}} \right) + {I_2}\left( {\sqrt {N + 3} } \right.\cdot n\cdot\left( {\left( {N + 3} \right)} \right.$ $\left. {\left. {\left. {\left. {\left( {n + 1} \right) + {n^2}} \right)} \right)} \right]} \right\}$, where $I_{outer}$ indicates the alternating iteration number of sub-problem 1 and sub-problem 2.

\section{Simulation Results}
In this section, simulation results are provided to verify the effectiveness of our proposed algorithm. The system is assumed to operate at 28 GHz carrier frequency. The complex gain ${\alpha _{vi}}\left( {{\alpha _{Ij}}} \right)$ of the channel model described in (1) and (8) is subject to ${\alpha _{vi}}\left( {{\alpha _{Ij}}} \right) \sim CN\left( {0,1} \right)$ distribution, and the number of paths and path loss exponent are set as 5 and 2, respectively \cite{Qinghua}. The angles of non-LoS (NLoS) links are uniform distribution for the given range \cite{add5}\cite{r2}. We assume that the location of CBS, SU and PU are at $(-80,29,15)$m, $(0,18.5,18.5)$m and $(80,29,15)$m, respectively. Without loss of generality, we assume that Eves are situated in circular regions centered at $(-44,25.5,18.5)$m,  $(16,28,18.5)$m, $(30,30,15)$m, $(-20,20,15)$m and $(50,20,30)$m. Based on this, we can attain the estimated AoD $\left( {\theta _k^c,\varphi _k^c} \right)$ at the center of Eves' uncertainty region. Then, for a given AoD uncertainty $\Delta$, we can obtain Eves' AoD upper and lower bound, i.e., $\left( {{\theta _{k,U}},{\varphi _{k,U}}} \right) = \left( {\theta _k^c + \frac{\Delta }{2},\varphi _k^c + \frac{\Delta }{2}} \right)$ and $\left( {{\theta _{k,L}},{\varphi _{k,L}}} \right) = \left( {\theta _k^c - \frac{\Delta }{2},\varphi _k^c - \frac{\Delta }{2}} \right)$. Consequently, we generate Eves' actual possible AoD value $\left( {\theta _k^{},\varphi _k^{}} \right) = \left( {{\theta _{k,L}}:\chi :{\theta _{k,U}},{\varphi _{k,L}}:\chi:{\varphi _{k,U}}} \right)$, where $\chi$ is taken as 0.1. The other parameters are set as $\sigma_P^2=-120$dBm, $K=2$, $\gamma_{th}=0$dB, $\varepsilon=10^{-3}$, unless otherwise stated. We also compare the results with other benchmarks: 1). the optimal solution with PCSI: the estimated CSI of the Eve channels is exactly the PCSI; 2). Non-robust scheme: it regards the estimated CSI of the Eve channels as PCSI; 3). Random IRS scheme: it only optimize IRS's phase shift matrix $\bf{Q}$ with random $\bf{w}$; 4). Random maximum ratio transmission (MRT) scheme: it performs the MRT at the CBS, i.e., ${\bf{w}} = \sqrt {P_c^{\max }} \frac{{{\bf{H}}_{CI}^H{{\bf{Q}}^H}{{\bf{h}}_{IS}}}}{{\left\| {{\bf{H}}_{CI}^H{{\bf{Q}}^H}{{\bf{h}}_{IS}}} \right\|}}$. For the random MRT scheme, the phase shift matrix $\bf{Q}$ of the IRS is randomly selected.

\begin{figure}[t]
	\centering
	\begin{minipage}[t]{0.24\textwidth}
		\centering
		\includegraphics[width=4.7cm]{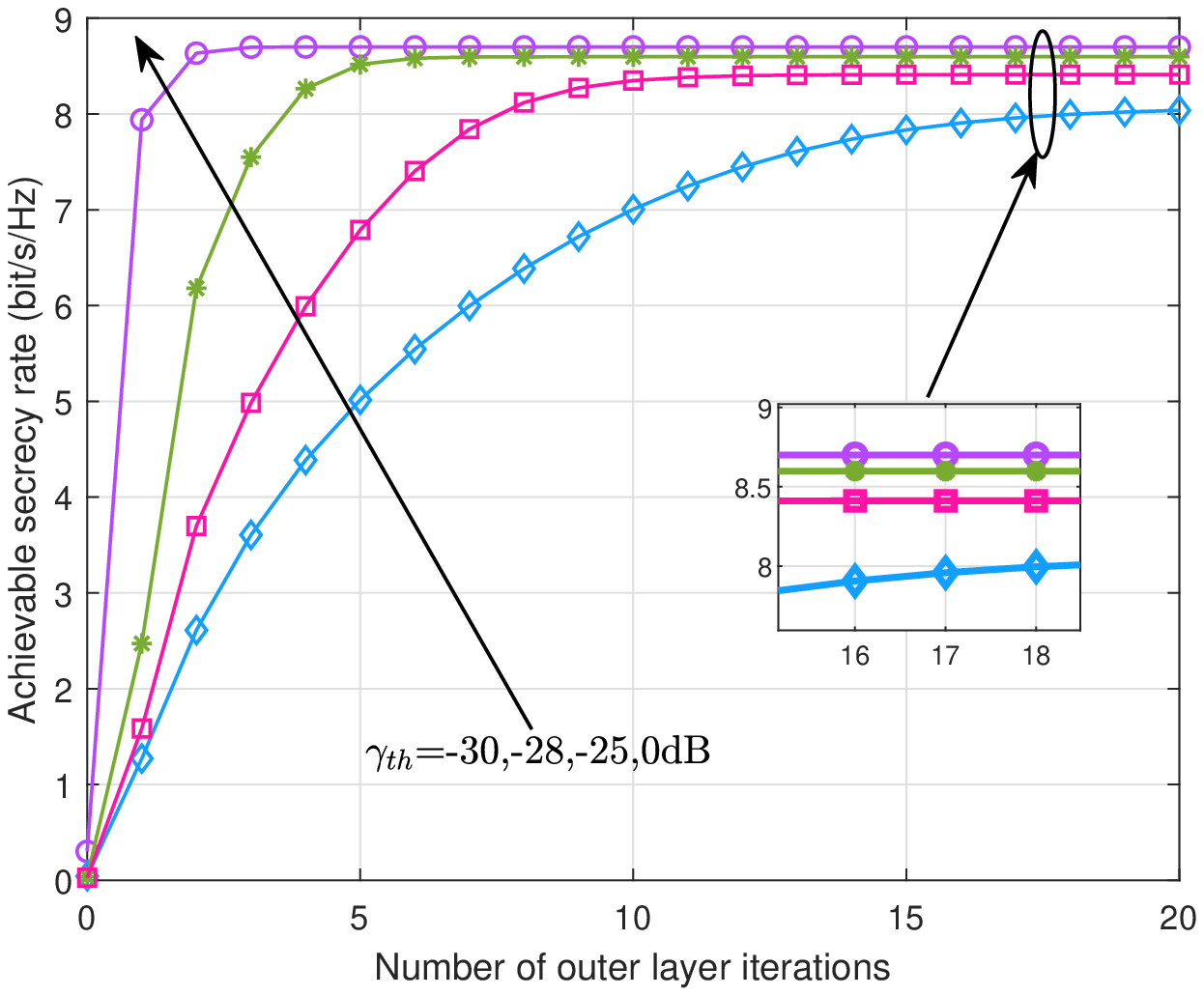}
		\caption{Convergence of proposed scheme with different IT thresholds.}
	\end{minipage}
	\begin{minipage}[t]{0.24\textwidth}
		\centering
		\includegraphics[width=4.7cm]{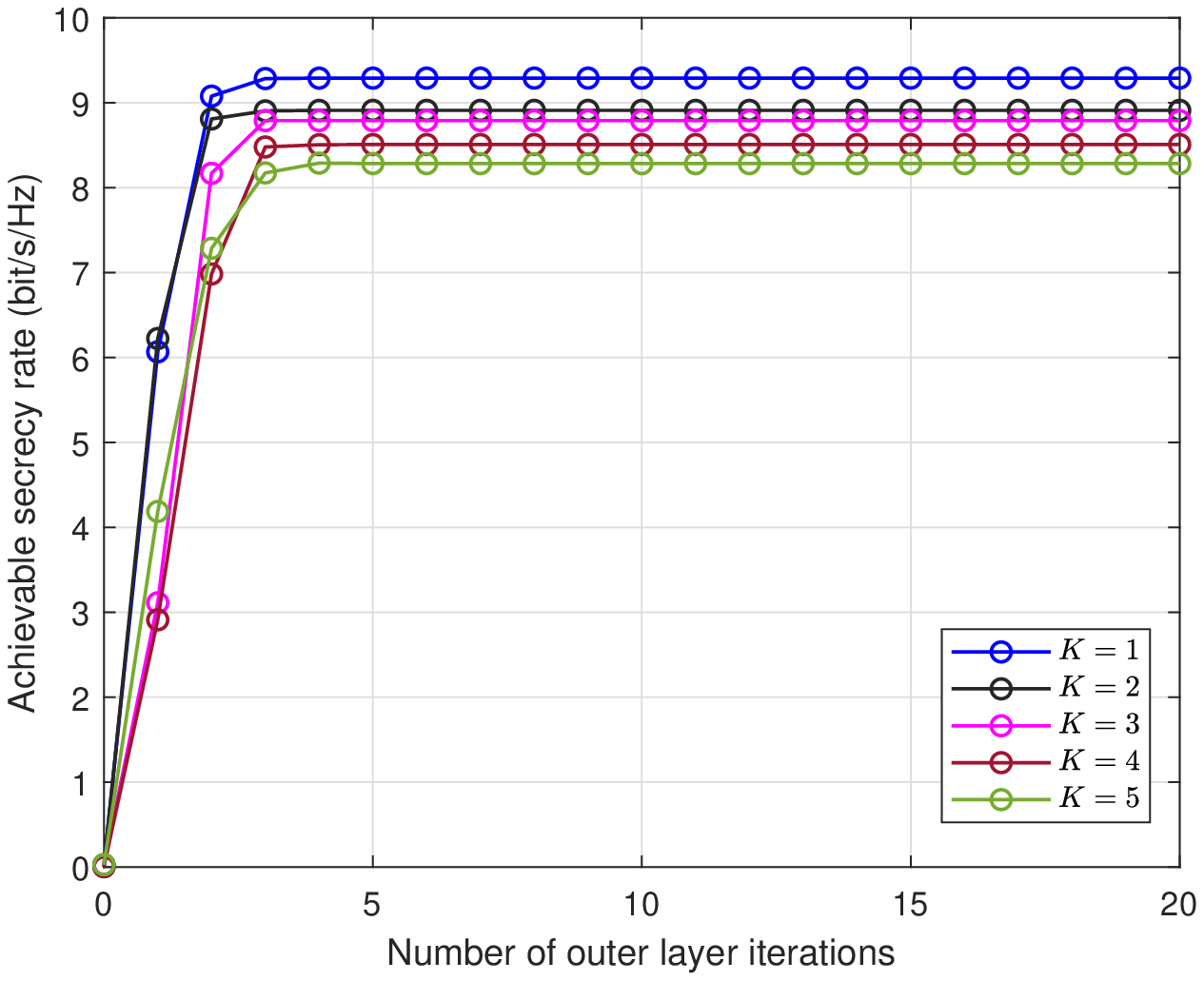}
		\caption{Convergence of proposed scheme with different numbers of Eve.}
	\end{minipage}
\end{figure}
Fig. 3 depicts the convergence of the robust secure BF scheme with different IT thresholds of PU, where $N=4\times4$, $M=8$, $\Delta=1^\circ$, $P_\text{c}^{\text{max}}=46$dBm. As we can see, the ASR of SU increases with the iteration number, and finally reaches a convergence within 20 iterations. Good ASR performance can be achieved with only 10 iteration rounds, validating the effectiveness of our proposed scheme. In addition, we can find that when $\gamma_{th}$ grows from -30dB to -28dB, the convergence ASR of SU increases significantly. The reason is that for a larger $\gamma_{th}$, PU has the ability to bear larger interference from CBS, which means that CBS can use more transmit power to increase SU's ASR. Moreover, it is interesting to find that when $\gamma_{th}$ further grows, i.e., from -28dB to -25dB and further to 0dB, there is only a slight increase in SU's ASR. This can be explained that when $\gamma_{th}$ is large enough, the IT constraint on PU tends to be relaxed. Furthermore, notice from the figure that the convergence speed becomes slower as $\gamma_{th}$ decreases. The reason is that when $\gamma_{th}$ becomes smaller, the IT constraint tends to be tight, making it difficult to find the optimal solution.

Fig. 4 shows the convergence of our proposed robust secure BF scheme with different numbers of Eve $K$, where $N=4\times4$, $M=8$, $\Delta=1^\circ$, $P_\text{c}^{\text{max}}=46$dBm. Since the constraints of (18) do not contradict each other, the constraints will not make (18) infeasible. Also, the constraints (18b)-(18d) are independent of Eves. Thus, increasing the number of Eves will not make the optimization problem (18) infeasible. As we can see from Fig. 4, the convergent ASR tends to decrease as the number of Eves increases. This is because that the increased number of Eves makes the cooperative Eves own more freedom to wiretap the signals, which can be reflected in (18a). In other words, a greater $K$ can increase the denominator of (18a), which further decreases the ASR. In addition, we can find that the convergence speed is almost the same under different numbers of Eve. This can be explained that the increase of $K$ only changes the number of $\sum\limits_{k = 1}^K {{\bf{H}}_k^H} {\bf{q}}{{\bf{q}}^H}{\bf{H}}_k^{}$ in (18a), which is equivalent to randomize the matrix value and thereby will not affect the convergence speed.

\begin{figure}[t]
	\centering
	\includegraphics[width=7.8cm]{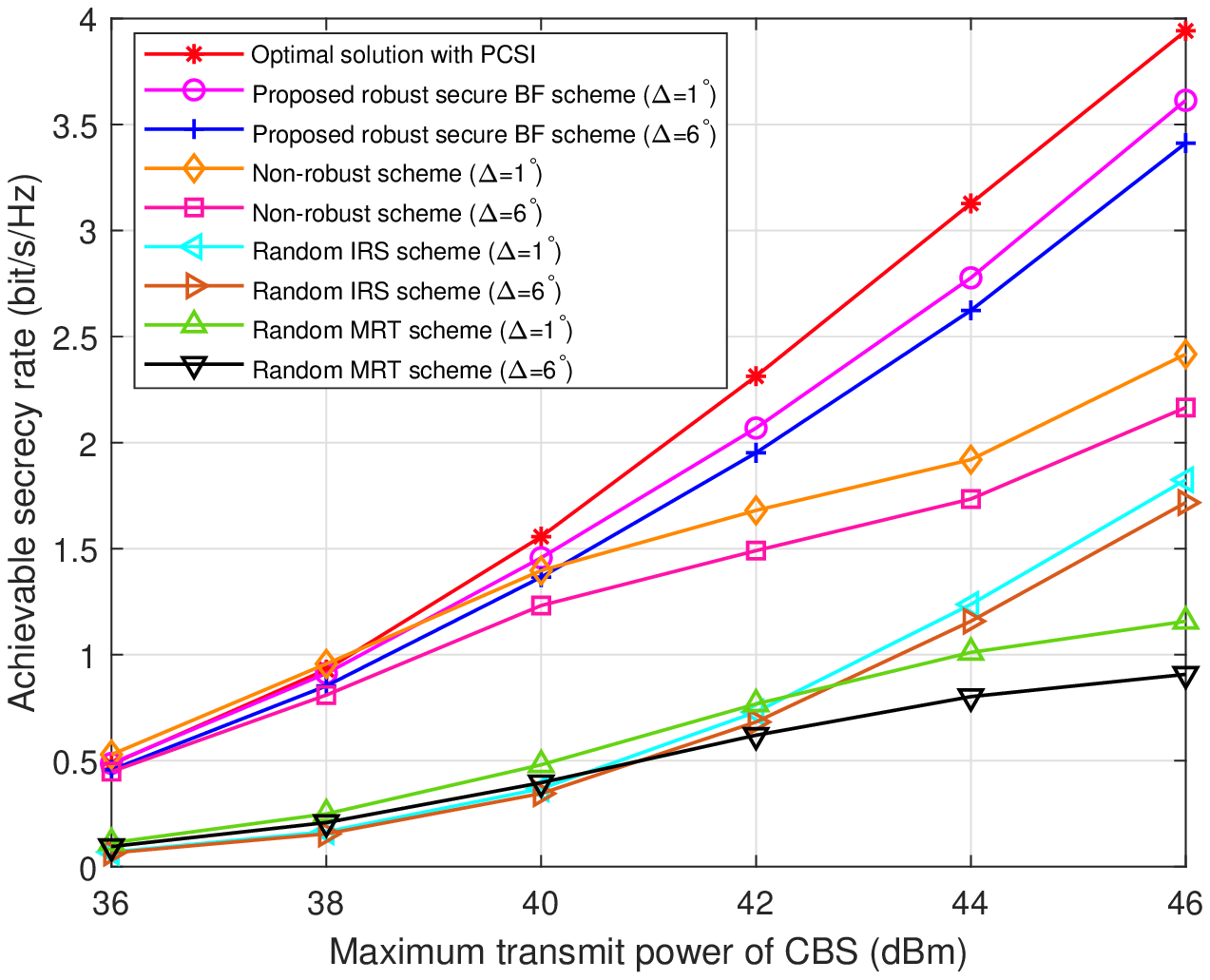}
	\caption{ASR versus $P_\text{c}^{\text{max}}$ with $N=2\times2$, $M=8$.}
	\label{fig1}
\end{figure}
\begin{figure}[t]
	\centering
	\includegraphics[width=7.8cm]{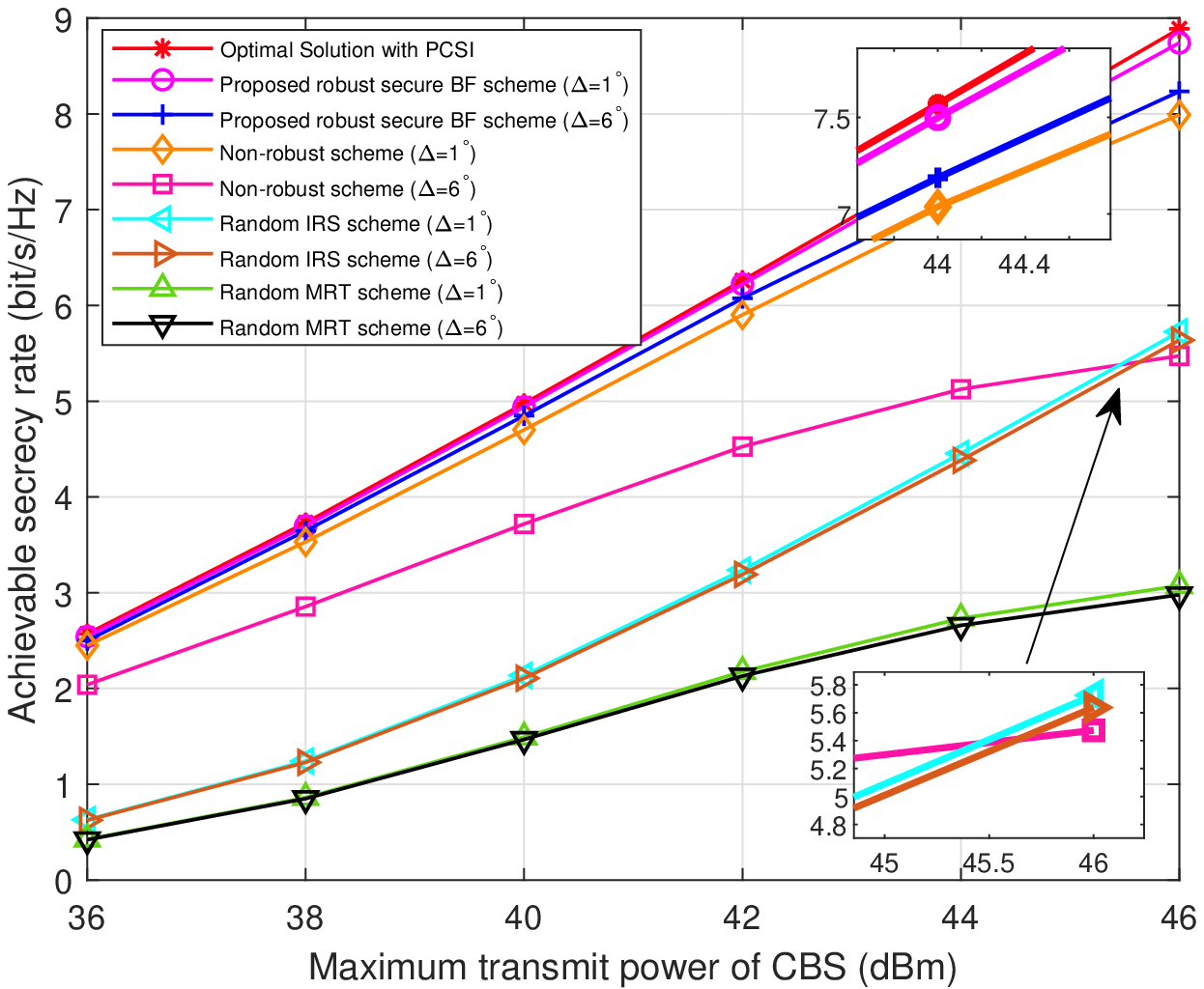}
	\caption{ASR versus $P_\text{c}^{\text{max}}$ with $N=4\times4$, $M=8$.}
	\label{fig1}
\end{figure}

\begin{figure}[t]
	\centering
	\begin{minipage}[t]{0.24\textwidth}
		\centering
		\includegraphics[width=4.7cm]{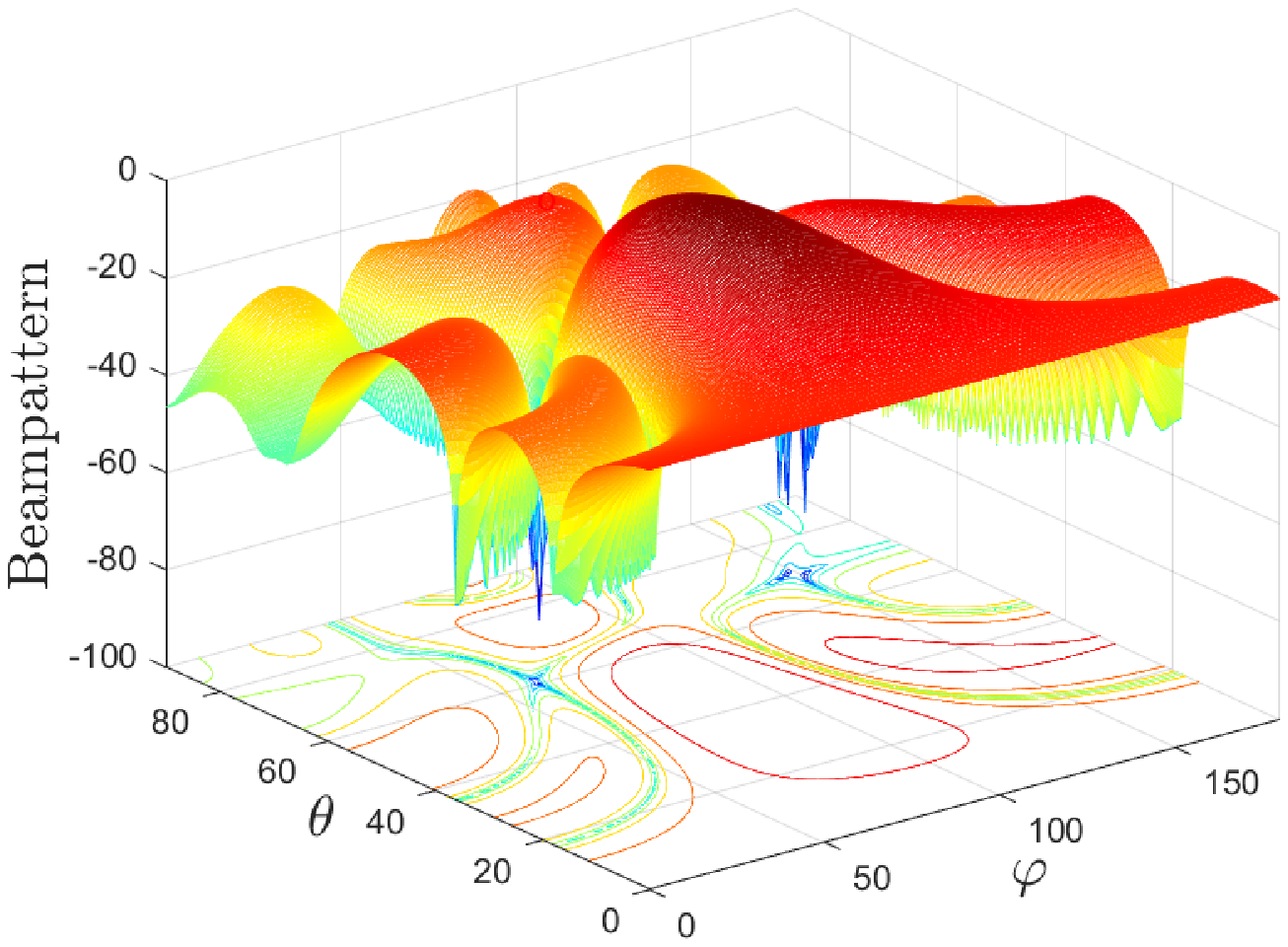}
		\caption{3D beampattern with $\Delta=1^\circ$.}
	\end{minipage}
	\begin{minipage}[t]{0.24\textwidth}
		\centering
		\includegraphics[width=4.7cm]{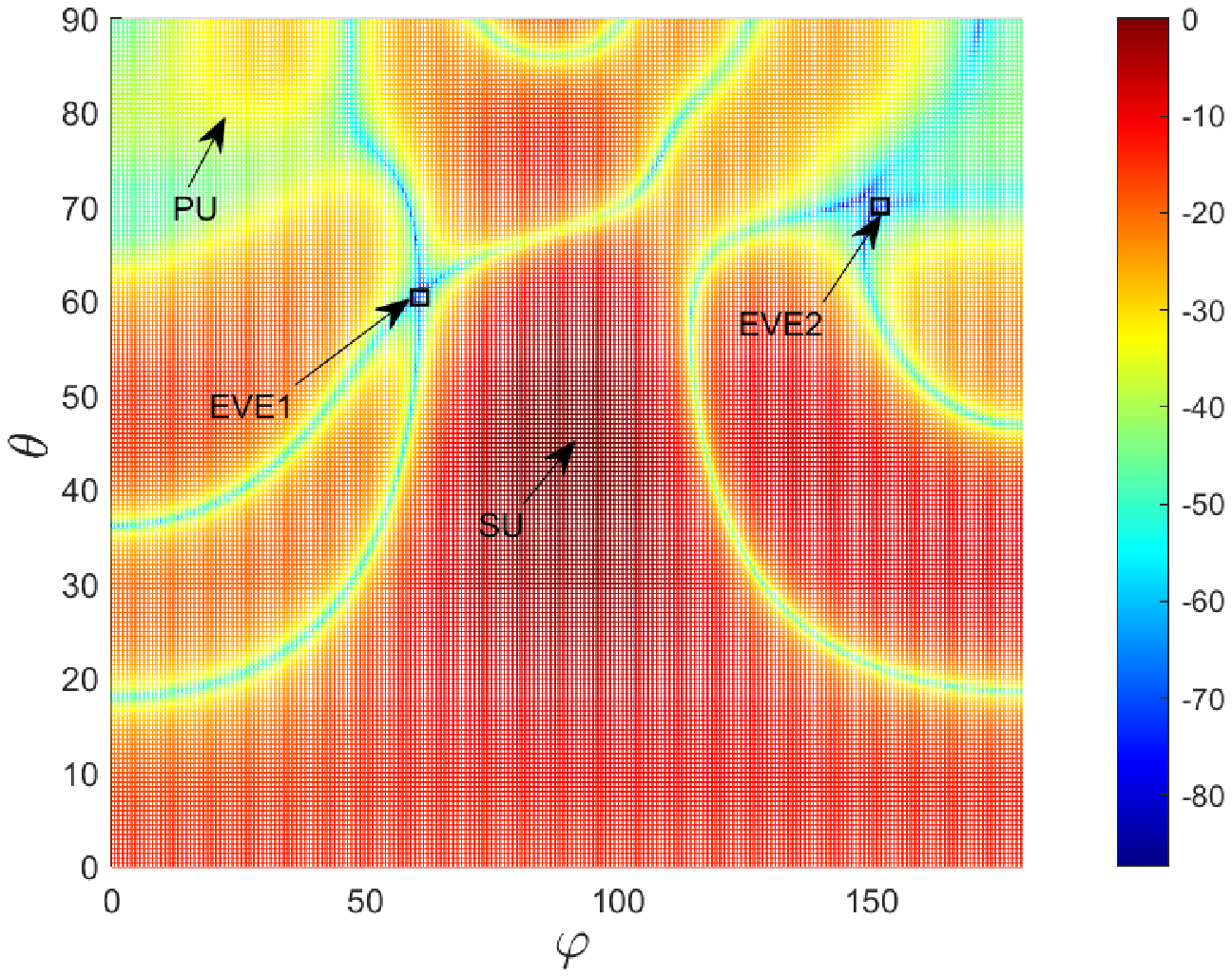}
		\caption{Beampattern from vertical vision with $\Delta=1^\circ$.}
	\end{minipage}
\end{figure}

Fig. 5 and Fig. 6 present the ASR of SU versus CBS's maximum transmit power with $N=2\times2$ and $N=4\times4$, respectively, where $M=8$. Since the optimal solution with PCSI is obtained under ideal condition, it is reasonable to regard it as the upper limit of ASR performance. As we can see, the ASRs of all schemes increase with the growth of the maximum transmit power of CBS and the ASR performance of all schemes under $\Delta=1^\circ$ is better than that under $\Delta=6^\circ$. Besides, the ASR of our proposed robust secure BF scheme is slightly worse than the optimal solution with PCSI, but much better than other benchmarks, verifying that the proposed scheme can effectively reduce the private signal leaked to Eves. As we can see from Fig. 5, the ASR performance of the non-robust scheme is better than that of the random IRS scheme when $N=2\times2$, which owes to the small number of IRS elements. Since the non-robust scheme optimizes both $\bf{w}$ and $\bf{q}$, and the random IRS only optimizes $\bf{q}$, the small number of IRS elements will make the random IRS scheme play less role. Under this circumstance, although the non-robust scheme ignores the difference between the estimated CSI and actual CSI of wiretap links, it can still maintain a relatively good ASR performance. However, the ASR performance of the non-robust scheme is worse than that of the random IRS scheme when $N=4\times4$, $P_\text{c}^{\text{max}}\ge45.5$dBm, $\Delta=6^\circ$, shown in Fig. 6. This is because that a greater number of IRS elements enables the random IRS scheme more effective than the non-robust scheme when the Eve channel uncertainty region is large, i.e., $\Delta=6^\circ$. It should be noted that this is true only when the maximum transmit power of CBS is large enough, i.e., $P_\text{c}^{\text{max}}\ge45.5$dBm, since the random IRS scheme ignores the optimization of $\bf{w}$ and only a large maximum transmit power of CBS can make the signal strength reflected by IRS acceptable. Furthermore, according to Fig. 6, we can find that when $N=2\times2$, $P_\text{c}^{\text{max}}\le42$dBm, both the random IRS scheme and random MRT scheme have poor ASR performance. This is because that as for the random IRS scheme, small value of the maximum transmit power of CBS $P_c^{max}$ will cause the signal reflected by IRS weak. In addition, the small number of IRS elements makes this scheme unable to fully play its role in optimizing $\bf{q}$. The above two reasons lead to poor ASR performance. As for the random MRT scheme, although it concentrates on the optimization of $\bf{w}$, small value of the maximum transmit power of CBS $P_c^{max}$ will generate weak signal, which is then reflected by IRS in a disorderly manner, thereby making the ASR performance poor. When $P_\text{c}^{\text{max}}$ is greater than 42dBm, the random IRS scheme has a significant advantage in ASR performance compared with the random MRT scheme, because $P_\text{c}^{\text{max}}$ is large enough to make the signal reflected by IRS of the random IRS scheme at an acceptable level. However, even a large enough $P_\text{c}^{\text{max}}$ can not improve the ASR performance of the random MRT scheme much, because the phase shift matrix $\bf{Q}$ that determines the received signal strength at the SU and Eves is not optimized. Moreover, we can see from Fig. 6 that the ASR performance of the random IRS scheme is always better than that of the random MRT scheme when $N=4\times4$, since the increasing number of IRS elements can make the random IRS scheme work well even with a small $P_\text{c}^{\text{max}}$, that is to say, the increasing number of IRS elements compensates for the decline of ASR performance resulting from a low value of $P_\text{c}^{\text{max}}$.

\begin{figure}[t]
	\centering
	\begin{minipage}[t]{0.24\textwidth}
		\centering
		\includegraphics[width=4.7cm]{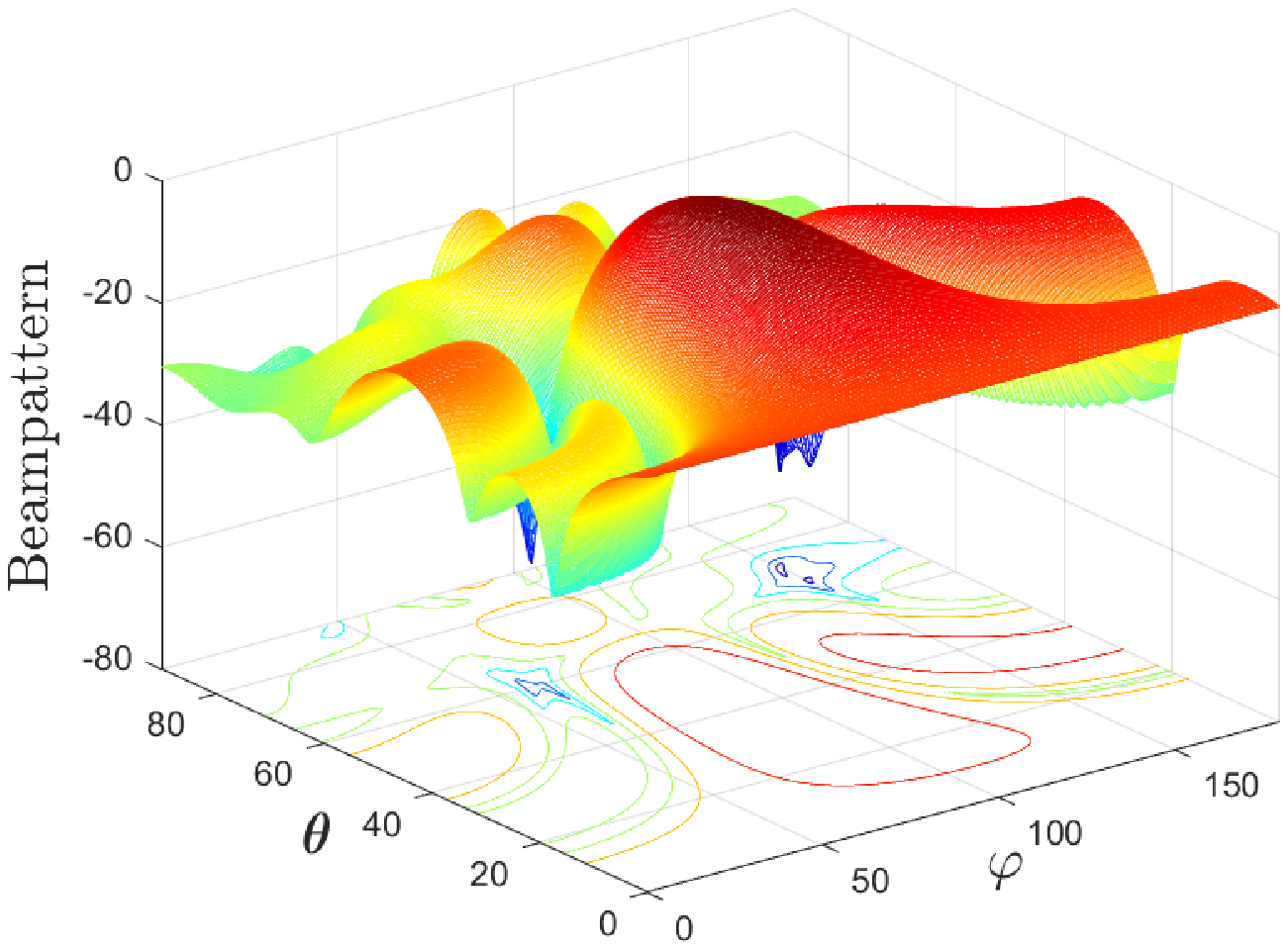}
		\caption{3D beampattern with $\Delta=6^\circ$.}
	\end{minipage}
	\begin{minipage}[t]{0.24\textwidth}
		\centering
		\includegraphics[width=4.7cm]{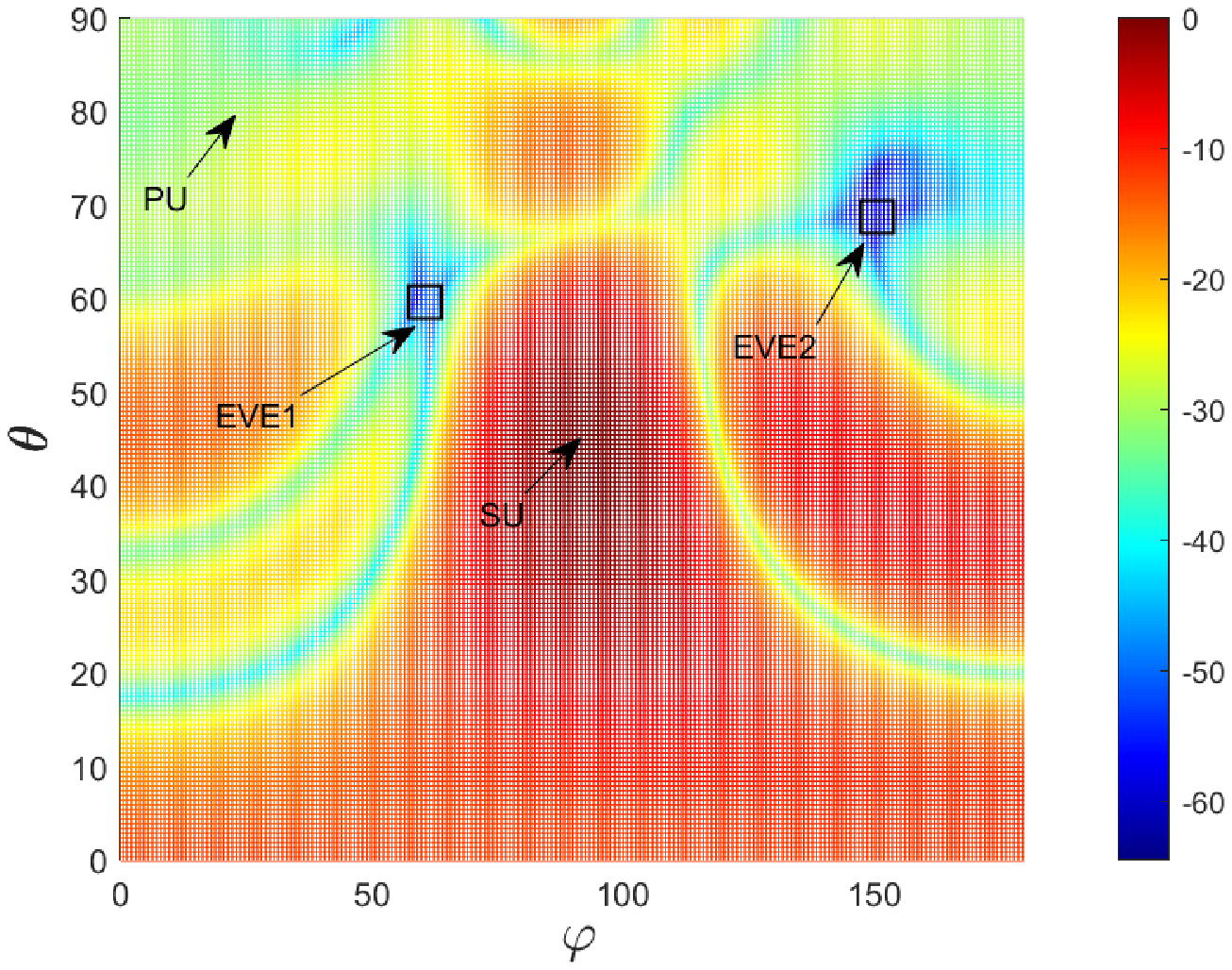}
		\caption{Beampattern from vertical vision with $\Delta=6^\circ$.}
	\end{minipage}
\end{figure}
Fig. 7 and Fig. 8 depict the 3D beampattern and corresponding vertical view of our proposed scheme with $N=6\times6$, $M=8$, $\Delta=1^\circ$, $P_\text{c}^{\text{max}}=46$dBm. As we can see, the received signal gain within Eves' uncertainty region is below -60dB (almost approaching 0), which is much lower than the 0dB of  SU. Meanwhile, the received signal gain at PU is around -40dB. The above observations strongly indicate that our proposed scheme can effectively enhance SU's signal strength, suppress the interference leakage on the PU, and simultaneously null the signal leaked to the Eves within the uncertainty region. The same analysis applies to Fig. 9 and Fig. 10. The difference is that the Eve uncertainty region of Fig. 9 and Fig. 10 is expanded to $\Delta=6^\circ$. Obviously, even if the Eve uncertainty region is expanded, our proposed scheme can still null the signal leaked to the Eves within the channel uncertainty region, which verifies the effectiveness of our proposed scheme for secure transmission even if there is a relatively big difference between the estimated CSI and actual CSI of wiretap links.

\begin{figure}[t]
	\centering
	\includegraphics[width=7.8cm]{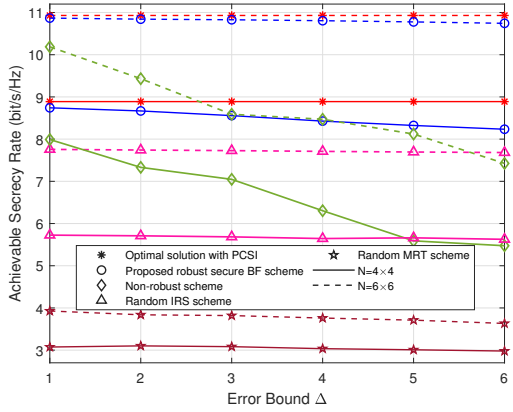}
	\caption{ASR versus AoD-based Eve error bound $\Delta$ with $M=8$.}
	\label{fig1}
\end{figure}
\begin{figure}[t]
	\centering
	\includegraphics[width=7.8cm]{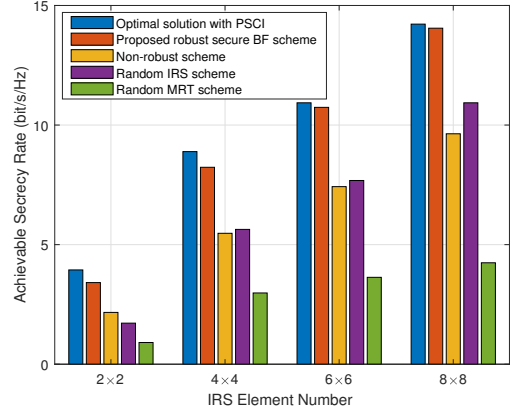}
	\caption{ASR versus IRS element number with $M=8$, $\Delta=6^\circ$, $P_\text{c}^{\text{max}}=46$dBm.}
	\label{fig1}
\end{figure}

Fig. 11 demonstrates the ASR versus Eves' AoD-based error bound $\Delta$ with $M=8$, $P_\text{c}^{\text{max}}=46$dBm. As expected, the optimal solution with PCSI and the proposed robust secure BF scheme outperform the other benchmarks in ASR performance. Compared with the non-robust scheme, the proposed robust secure BF scheme is less susceptible to the change of channel uncertainties due to the robust design based on the Eves' channel uncertainty region, which can also account for random IRS scheme's insensitivity to the channel uncertainty. However, the random IRS scheme ignores the transmit BF design at the CBS, causing the ASR performance worse than that of our proposed robust secure BF scheme. Moreover, exploiting the channel only related to SU makes the ASR performance of random MRT scheme almost not influenced by the Eve channel uncertainties, and much lower than other benchmarks.

Fig. 12 depicts the ASR performance versus IRS element number with $M=8$, $\Delta=6^\circ$, $P_\text{c}^{\text{max}}=46$dBm. We can find that better ASR performance can be achieved with a greater number of IRS elements for all schemes, which is particularly obvious for the increase from $2\times2$ to $4\times4$. Besides, when the number of IRS elements becomes larger, the ASR performance of our proposed scheme is closer to that of the optimal solution with PCSI, indicating that the more the number of IRS elements, the better the IRS can reflect the signal in the direction of SU and away from the Eve estimation region. In addition, we can find that only when $N=2\times2$, the ASR performance of the non-robust scheme is better than that of the random IRS scheme. As $N$ is greater than $2\times2$, the random IRS scheme 's ASR performance outperforms the non-robust scheme's. This can be explained that the increasing number of IRS elements enables the random IRS scheme more effective than the non-robust scheme when the Eve channel uncertainty region is large, which corresponds with the results in Fig. 5 and Fig. 6. As for the random MRT scheme, since it does not take the optimization of IRS phase shift matrix $\bf{Q}$ into account, the ASR performance always keeps at a low level. For this scheme, even if the value of $N$ is large, due to the unoptimized IRS phase shift matrix $\bf{Q}$, the IRS only reflects the signal in a disorderly manner, which has no substantial help to the performance improvement. 

\begin{figure}[t]
	\centering
	\includegraphics[width=7.8cm]{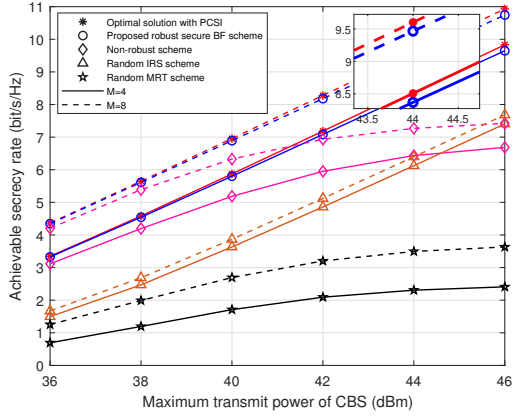}
	\caption{ASR versus $P_\text{c}^{\text{max}}$ with different antenna numbers of CBS.}
	\label{fig1}
\end{figure}
Fig. 13 shows the ASR performance versus CBS's maximum transmit power with different antenna numbers of CBS, where $N=6\times6$, $\Delta=6^\circ$. As we can see, all schemes can achieve a better ASR performance with a greater number of CBS antennas, especially for the optimal solution with PCSI, proposed robust secure BF scheme, non-robust scheme and random MRT scheme. However, the performance improvement of the random IRS scheme is not very obvious. This is due to the fact that the random IRS scheme adopts random transmit BF $\bf{w}$, resulting in that the increasing number of CBS antenna does not contribute much to the performance improvement. As for other schemes, the larger the number of CBS antennas, the greater the signal strength reflected by IRS, which further improves the ASR performance.

\begin{figure}[t]
	\centering
	\includegraphics[width=7.8cm]{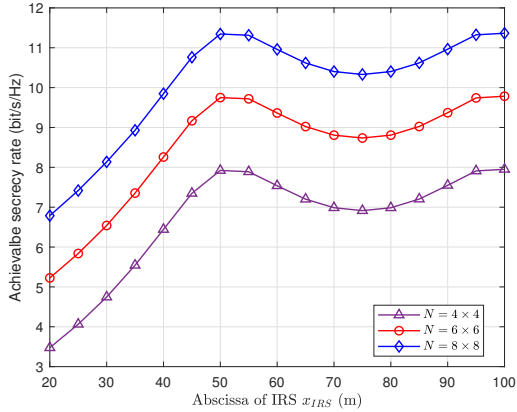}
	\caption{ASR versus the abscissa of IRS $x_{IRS}$ with $y_{IRS}=10m$, $M=8$, $P_c^{\max } = 46$dBm, $\Delta  = 1^\circ $.}
	\label{fig1}
\end{figure}

\begin{figure}[t]
	\centering
	\includegraphics[width=7.8cm]{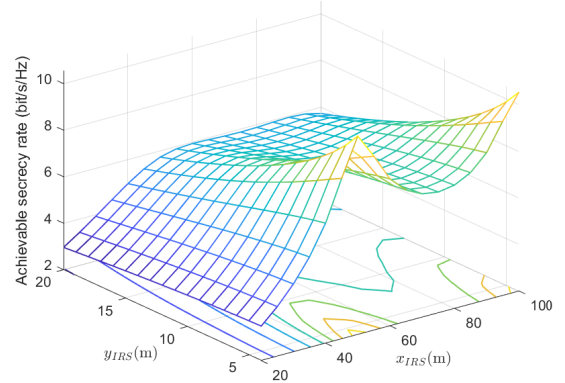}
	\caption{ASR versus the position of IRS in a two-dimensional space with $N=4\times4$, $M=8$, $P_c^{\max } = 46$dBm, $\Delta  = 1^\circ $.}
	\label{fig1}
\end{figure}

In order to vividly investigate the impact of the position of IRS, we consider a simulation scenario in a two-dimensional space: the PU, CBS, IRS and SU are located at (0, 0), (50, 0), ($x_{IRS}$, $y_{IRS}$) and (100, 0) in meters, respectively. Also, the two Eves are situated in circular regions centered at (80, 0) and (90, 0) in meters. Fig. 14 depicts ASR versus the abscissa of IRS $x_{IRS}$ with $y_{IRS}=10m$, $M=8$, $P_c^{\max } = 46$dBm, $\Delta  = 1^\circ $. Fig. 15 shows ASR versus the position of IRS in a two-dimensional space with $N=4\times4$, $M=8$, $P_c^{\max } = 46$dBm, $\Delta  = 1^\circ $. From these figures, we can see that for a given $y_{IRS}$, there are two optimal locations of $x_{IRS}$, i.e., $x_{IRS} = 50m$ and $x_{IRS} = 100m$. Furthermore, the IRS should avoid being deployed in the middle location between the CBS and SU. Actually, the worst case occurs when the IRS is located far away from SU and close to the PU. These results indicate that, in order to obtain a higher ASR, the IRS should be deployed in the vicinity of CBS or SU. In addition, from Fig. 15, we can find that when the vertical height of the IRS $y_{IRS}$ is getting higher and higher, the variation of the curve between CBS and SU becomes flattened. This can be explained that the higher vertical height of IRS $y_{IRS}$ makes the variation in the horizontal coordinate of IRS $x_{IRS}$ appear smaller.

\begin{figure}[t]
	\centering
	\includegraphics[width=7.8cm]{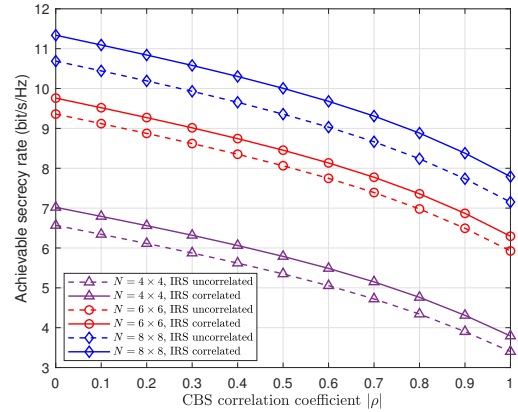}
	\caption{ASR versus CBS spatial correlation $\left| \rho  \right|$ with $M=8$, $P_c^{\max } = 46$dBm, $\Delta  = 1^\circ $.}
	\label{fig1}
\end{figure}

To investigate the impact of spatial correlation, we introduce the adopted spatial correlation models for CBS and IRS. 1) Spatial correlation at CBS: We apply the Kronecker correlation model at the CBS, which is suitable for uniform linear arrays, such that the $ij$-th entry of ${{{\bf{R}}_{CBS}}}$ is given as
\begin{align}
		{\left[ {{{\bf{R}}_{CBS}}} \right]_{i,j}} = \left\{ \begin{array}{l}
			{\rho ^{\left( {j - i} \right)}},\quad{\rm{      if  }}\quad i \le j\\
			{\left( {{\rho ^{\left| {j - i} \right|}}} \right)^*}{\rm{,  otherwise}}
		\end{array} \right.{\rm{    }}\forall \left\{ {i,j} \right\} \le \left\{ {1, \cdots ,M} \right\},\notag
\end{align}
where $\rho  \in \mathbb{C}$ is the correlation coefficient satisfying $\left| \rho  \right| \le 1$. 2) Spatial correlation at IRS: For IRS, we utilize the spatial correlation proposed by \cite{correlation} for rectangular surface. In particular, the $mn$-th entry of $\bf{R}$ is given as
\begin{align}
		{\left[ {\bf{R}}_{IRS} \right]_{m,n}} = {\rm{sinc}}\left( {\frac{{2\left\| {{u_m} - {u_n}} \right\|}}{\lambda }} \right),\forall \left\{ {m,n} \right\} \in \left\{ {1,2, \cdots ,N} \right\},\notag
\end{align}
where $\left\| {{u_m} - {u_n}} \right\|$ is the distance between $m$-th and $n$-th reflecting elements. $\lambda$ is the wavelength. ${\rm{sinc}}\left( x \right) = {{\sin \left( {\pi x} \right)} \mathord{\left/{\vphantom {{\sin \left( {\pi x} \right)} {\left( {\pi x} \right)}}} \right.\kern-\nulldelimiterspace} {\left( {\pi x} \right)}}$ is the sinc function. As such, the channel matrix of the CBS-IRS link ${{\bf{H}}_{CI}}$ and the channel vector of the IRS-user link ${{\bf{h}}_{Iv}}$ are rewritten as ${{\bf{\tilde H}}_{CI}} = {\bf{R}}_{CBS}^{\frac{1}{2}}{{\bf{H}}_{CI}}{\bf{R}}_{IRS}^{\frac{1}{2}}$ and ${{{\bf{\tilde h}}}_{Iv}} = {\bf{R}}_{IRS}^{\frac{1}{2}}{{\bf{h}}_{Iv}},v \in \left\{ {S,k,P} \right\}$, respectively. Based on the spatial correlation models above, Fig. 16 illustrates the ASR versus CBS spatial correlation $\left| \rho  \right|$ under uncorrelated and correlated IRS conditions with $M=8$, $P_c^{\max } = 46$dBm, $\Delta  = 1^\circ $. As we can see, with the increase of CBS spatial correlation $\left| \rho  \right|$, the ASR decreases. We consider the reason to be that a higher channel correlation leads to a smaller solution space, and thereby reduces the ASR as a result. In addition, we can find that the spatial correlation at IRS offers improvement in the ASR performance. Thus, the results in Fig. 16 clearly show that higher spatial correlation at CBS is not desired, while the opposite holds for the correlation at IRS.

\section{Conclusion}
This paper has investigated the robust secure BF design in the IRS-assisted mmWave CRNs. Specifically, by using the AoD-based wiretap channel uncertainty, we formulate a problem to make SU's worst-case ASR maximum under the premise that CBS's transmit power, PU's IT and IRS's unit modulus meet respective constraints. To solve the non-convex problem with coupled variables, an efficient alternating optimization algorithm is proposed. Finally, simulation results indicate that the ASR performance of our proposed scheme has a small gap with that of the optimal solution with PCSI compared with the other benchmarks. Furthermore, our proposed robust secure BF scheme can still null the signal leaked to the Eves within a large channel uncertainty region, which validates the effectiveness of our proposed robust secure BF scheme even if there is a relatively big difference between the estimated CSI and actual CSI of wiretap links. In addition, with the increase of the number of IRS, our proposed robust secure BF scheme can obtain better ASR performance, which is closer to that of the optimal solution with PCSI. Our proposed algorithm can be extended to the case with multiple SUs and PUs when the CBS transmits the same signal to the SUs. However, when the CBS transmits different signals to corresponding SU, there will exist inter-user interference, which will be the content of our follow-up research. In addition, we are looking forward to investigating the low-complexity alternating direction method of multipliers (ADMM) algorithm \cite{pan} to solve the phase shift optimization problem in our follow-up work.

\appendices

\section{Proof of Proposition 1}
Since ${\Psi _k}$ is the convex hull of ${\Lambda _k}$, i.e., ${\Lambda _k} \subseteq {\Psi _k}$, we can attain $\mathop {\max }\limits_{{{\bf{F}}_k} \in {\Lambda _k}} {{\bf{w}}^H}{\bf{H}}_{CI}^H\sum\limits_{k = 1}^K {{{\bf{F}}_k}{{\bf{H}}_{CI}}} {\bf{w}} \le \mathop {\max }\limits_{{{\bf{F}}_k} \in {\Psi _k}} {{\bf{w}}^H}{\bf{H}}_{CI}^H\sum\limits_{k = 1}^K {{{\bf{F}}_k}{{\bf{H}}_{CI}}} {\bf{w}}$, which further leads to
\setcounter{equation}{45}
\begin{equation}
	\begin{split}
\begin{array}{l}
	\mathop {\min }\limits_{{{\bf{F}}_k} \in {\Lambda _k}} \frac{{{{\bf{w}}^H}{\bf{H}}_S^H{\bf{q}}{{\bf{q}}^H}{{\bf{H}}_S}{\bf{w}} + \sigma _S^2}}{{{{\bf{w}}^H}{\bf{H}}_{CI}^H\sum\limits_{k = 1}^K {{{\bf{F}}_k}{{\bf{H}}_{CI}}} {\bf{w}} + \sigma ^2}}
	\ge \mathop {\min }\limits_{{{\bf{F}}_k} \in {\Psi _k}} \frac{{{{\bf{w}}^H}{\bf{H}}_S^H{\bf{q}}{{\bf{q}}^H}{{\bf{H}}_S}{\bf{w}} + \sigma _S^2}}{{{{\bf{w}}^H}{\bf{H}}_{CI}^H\sum\limits_{k = 1}^K {{{\bf{F}}_k}{{\bf{H}}_{CI}}} {\bf{w}} + \sigma ^2}}.
\end{array}
\end{split}
\end{equation}

On the other hand, according to convex hull's feature \cite{49} and the formula (20), we can decompose ${{\bf{w}}^H}{\bf{H}}_{CI}^H{{\bf{F}}_k}{{\bf{H}}_{CI}}{\bf{w}}\left( {\forall {{\bf{F}}_k} \in {\Psi _k}} \right)$ as
\begin{equation}
\begin{array}{l}
	\quad{{\bf{w}}^H}{\bf{H}}_{CI}^H{{\bf{F}}_k}{{\bf{H}}_{CI}}{\bf{w}}\\
	= {{\bf{w}}^H}{\bf{H}}_{CI}^H\left( {{\mu _{k,1}}{{\bf{F}}_{k,1}} +  \cdots  + {\mu _{k,{M_k}}}{{\bf{F}}_{k,{M_k}}}} \right){{\bf{H}}_{CI}}{\bf{w}}\\
	= {\mu _{k,1}}{{\bf{w}}^H}{\bf{H}}_{CI}^H{{\bf{F}}_{k,1}}{{\bf{H}}_{CI}}{\bf{w}} +  \cdots {\mu _{k,{M_k}}}{{\bf{w}}^H}{\bf{H}}_{CI}^H{{\bf{F}}_{k,{M_k}}}{{\bf{H}}_{CI}}{\bf{w}}.
\end{array}
\end{equation}Due to the fact that $\forall {{\bf{F}}_k} \in {\Psi _k} = \left\{ {\sum\limits_{i = 1}^{{M_k}} {{\mu _{k,i}}{{\bf{F}}_{k,i}}} {\rm{|}}\sum\limits_{i = 1}^{{M_k}} {{\mu _{k,i}} = 1,\;{\mu _{k,i}} \ge 0} } \right\}$, there must exist a ${{\bf{F}}_{k,m}} \in {\Lambda _k}$ which satisfies 
${{\bf{w}}^H}{\bf{H}}_{CI}^H{{\bf{F}}_{k,m}}{{\bf{H}}_{CI}}{\bf{w}} \ge {{\bf{w}}^H}{\bf{H}}_{CI}^H{{\bf{F}}_k}{{\bf{H}}_{CI}}{\bf{w}}$. Then, one can further obtain that there must exist a ${{\bf{F}}_{k,m}} \in {\Lambda _k}$ meeting the condition
\begin{equation}
	\begin{array}{l}
\frac{{{{\bf{w}}^H}{\bf{H}}_S^H{\bf{q}}{{\bf{q}}^H}{{\bf{H}}_S}{\bf{w}} + \sigma _S^2}}{{{{\bf{w}}^H}{\bf{H}}_{CI}^H{{\bf{F}}_{k,m}}{{\bf{H}}_{CI}}{\bf{w}} + \sigma^2}} \le \frac{{{{\bf{w}}^H}{\bf{H}}_S^H{\bf{q}}{{\bf{q}}^H}{{\bf{H}}_S}{\bf{w}} + \sigma _S^2}}{{{{\bf{w}}^H}{\bf{H}}_{CI}^H{{\bf{F}}_k}{{\bf{H}}_{CI}}{\bf{w}} + \sigma^2}},
\end{array}
\end{equation}Or to put it another way,
\begin{equation}
\begin{array}{l}
	\quad\mathop {\min }\limits_{{{\bf{F}}_{k,m}} \in {\Lambda _k}} \frac{{{{\bf{w}}^H}{\bf{H}}_S^H{\bf{q}}{{\bf{q}}^H}{{\bf{H}}_S}{\bf{w}} + \sigma _S^2}}{{{{\bf{w}}^H}{\bf{H}}_{CI}^H\sum\limits_{k = 1}^K {{{\bf{F}}_{k,m}}{{\bf{H}}_{CI}}} {\bf{w}} + \sigma ^2}}\\
	\le \mathop {\min }\limits_{{{\bf{F}}_k} \in {\Psi _k}} \frac{{{{\bf{w}}^H}{\bf{H}}_S^H{\bf{q}}{{\bf{q}}^H}{{\bf{H}}_S}{\bf{w}} + \sigma _S^2}}{{{{\bf{w}}^H}{\bf{H}}_{CI}^H\sum\limits_{k = 1}^K {{{\bf{F}}_k}{{\bf{H}}_{CI}}} {\bf{w}} + \sigma ^2}}.
\end{array}
\end{equation}

Therefore, by combining the formula (46) and the formula (49), we can complete the proof.

\section{Proof of Proposition 2}
First of all, let us define a function $g\left( {{\bf{w}},{{\bf{F}}_k}} \right)$, given by
\begin{equation}
\begin{array}{c}
	g\left( {{\bf{w}},{{\bf{F}}_k}} \right) = \frac{{{{\bf{w}}^H}{\bf{H}}_S^H{\bf{q}}{{\bf{q}}^H}{{\bf{H}}_S}{\bf{w}} + \sigma _S^2}}{{{{\bf{w}}^H}{\bf{H}}_{CI}^H\sum\limits_{k = 1}^K {{{\bf{F}}_k}{{\bf{H}}_{CI}}} {\bf{w}} + \sigma^2}}\\
	\quad\quad\quad\quad\quad\buildrel \Delta \over = \frac{{{\rm{tr}}\left( {{\bf{H}}_S^H{\bf{q}}{{\bf{q}}^H}{{\bf{H}}_S}{\bf{W}}} \right) + \sigma _S^2}}{{{\rm{tr}}\left( {{\bf{H}}_{CI}^H\sum\limits_{k = 1}^K {{{\bf{F}}_k}{{\bf{H}}_{CI}}{\bf{W}}} } \right) + \sigma ^2}},
\end{array}
\end{equation}
where ${\bf{W}} = {{\bf{w}}^H}{\bf{w}}$. As a result of the convex set of ${\Psi _k}$, $g\left( {{\bf{w}},{{\bf{F}}_k}} \right)$ is a  convex function with respect to ${{\bf{F}}_k}$ over ${\Psi _k}$ for any $\bf{W}$. Referring to \textit{Theorem 2.1} in \cite{54}, there must exist a saddle point $\left( {{{\bf{W}}^*},{\bf{F}}_k^*} \right)$ which satisfies
$g\left( {{\bf{W}},{\bf{F}}_k^*} \right) \le g\left( {{{\bf{W}}^*},{\bf{F}}_k^*} \right) \le g\left( {{{\bf{W}}^*},{\bf{F}}_k^{}} \right),\forall {\bf{W}},\forall {{\bf{F}}_k} \in {\Psi _k}$. By exploiting the property of the saddle point in the max-min problem \cite{55}, we can obtain
\begin{equation}
	\begin{split}
g\left( {{{\bf{W}}^*},{\bf{F}}_k^*} \right) &= \mathop {\max }\limits_{\bf{W}} \mathop {\min }\limits_{{{\bf{F}}_k} \in {\Psi _k}} g\left( {{\bf{W}},{\bf{F}}_k^{}} \right) \\
&= \mathop {\min }\limits_{{{\bf{F}}_k} \in {\Psi _k}} \mathop {\max }\limits_{\bf{W}} g\left( {{\bf{W}},{\bf{F}}_k^{}} \right),
	\end{split}
\end{equation}
which indicates that the max-min problem is equivalent with the min-max problem, and they have the same solution at the saddle point $\left( {{{\bf{W}}^*},{\bf{F}}_k^*} \right)$.
Thus, one can obtain
\begin{equation}
\begin{array}{l}
\quad	\mathop {\max }\limits_{\bf{W}} \mathop {\min }\limits_{{{\bf{F}}_k} \in {\Psi _k}} \frac{{{\rm{tr}}\left( {{\bf{H}}_S^H{\bf{q}}{{\bf{q}}^H}{{\bf{H}}_S}{\bf{W}}} \right) + \sigma _S^2}}{{{\rm{tr}}\left( {{\bf{H}}_{CI}^H\sum\limits_{k = 1}^K {{{\bf{F}}_k}{{\bf{H}}_{CI}}{\bf{W}}} } \right) + \sigma ^2}}\\
	= \mathop {\min }\limits_{{{\bf{F}}_k} \in {\Psi _k}} \mathop {\max }\limits_{\bf{W}} \frac{{{\rm{tr}}\left( {{\bf{H}}_S^H{\bf{q}}{{\bf{q}}^H}{{\bf{H}}_S}{\bf{W}}} \right) + \sigma _S^2}}{{{\rm{tr}}\left( {{\bf{H}}_{CI}^H\sum\limits_{k = 1}^K {{{\bf{F}}_k}{{\bf{H}}_{CI}}{\bf{W}}} } \right) + \sigma ^2}},
\end{array}
\end{equation}
which can be further written in the form of $\bf{w}$. Thus, we have proved \textit{Proposition 2}.

\section{Proof of Proposition 3}
Rewrite the problem (36) as
\begin{align}
	\mathop {\min }\limits_{\bf{\Theta }} \mathop {\max }\limits_{{{\bf{G}}_k} \in {\Upsilon _k}} &{\rm{ }}\frac{{\sum\limits_{k = 1}^K {{\rm{tr}}\left( {{{\bf{G}}_k}{\bf{\Theta }}} \right)}  + \sigma _{}^2}}{{{\rm{tr}}\left( {{{\bf{H}}_S}{\bf{w}}{{\bf{w}}^H}{\bf{H}}_S^H{\bf{\Theta }}} \right) + \sigma _S^2}}\tag{53a}\\
	{\text{ s.t.}}\quad
&{\rm{tr}}\left( {{{\bf{H}}_P}{\bf{w}}{{\bf{w}}^H}{\bf{H}}_P^H{\bf{\Theta }}} \right) \le I_p^{th}, \tag{53b}\\
&{\left[ {\bf{\Theta }} \right]_{n,n}} = 1, \forall n \in \left\{ {1, \cdots ,N} \right\}=\mathbb N, \tag{53c}\\
&\rm{rank}\left( {\bf{\Theta }} \right) = 1\tag{53d}.
\end{align}
With the help of the auxiliary variable ${\bf{R}}=r{\bf{\Theta}}$, $r>0$, (53) can be further reformulated as
\begin{align}
	\mathop {\min }\limits_{{\bf{R }}{\underline\succ}0, r\ge0} \mathop {\max }\limits_{{{\bf{G}}_k} \in {\Upsilon _k}} &{\rm{ }}\frac{{\sum\limits_{k = 1}^K {{\rm{tr}}\left( {{{\bf{G}}_k}{\bf{R }}} \right)}  + r\sigma _{}^2}}{{{\rm{tr}}\left( {{{\bf{H}}_S}{\bf{w}}{{\bf{w}}^H}{\bf{H}}_S^H{\bf{R }}} \right) + r\sigma _S^2}}\tag{54a}\\
	{\text{ s.t.}}\quad
	&{\rm{tr}}\left( {{\bf{R }}{{\bf{H}}_P}{\bf{WH}}_P^H} \right) \le rI_p^{th}, \tag{54b}\\
	&{\rm{rank}}({\bf{R}})=1,\quad{\rm{  and  }}\quad{\left[ {\bf{R }} \right]_{n, n}} = r, \forall n \in {\mathbb N},\tag{54c}
\end{align}
which can be equivalently written as
\begin{align}
\mathop {\min }\limits_{{\bf{R}}{\underline\succ}0,r \ge 0} \mathop {\max }\limits_{{{\bf{G}}_k} \in {\Upsilon _k}} &\sum\limits_{k = 1}^K {{\rm{tr}}\left( {{{\bf{G}}_k}{\bf{R}}} \right)}  + r\sigma _{}^2\tag{55a}\\
	{{\rm{ s}}{\rm{.t}}{\rm{.}}\quad }&{\rm{tr}}\left( {{{\bf{H}}_S}{\bf{w}}{{\bf{w}}^H}{\bf{H}}_S^H{\bf{R}}} \right) + r\sigma _S^2 = 1,\tag{55b}\\
	&(54b)-(54c). \tag{55c}
\end{align}
Rewriting (55b) as ${\rm{tr}}\left( {{{\bf{H}}_S}{\bf{w}}{{\bf{w}}^H}{\bf{H}}_S^H{\bf{R}}} \right) + r\sigma _S^2 \ge 1$ does not change the optimal solution of (55), which can be explained as follows: assume that $\left( {{{\bf{R}}^*},{r^*}} \right)$ is the optimal solution satisfying ${\rm{tr}}\left( {{{\bf{H}}_S}{\bf{w}}{{\bf{w}}^H}{\bf{H}}_S^H{\bf{R}}} \right) + r\sigma _S^2 > 1$. Then, there definitely exists a certain vale $0<\beta<1$, enabling us to choose a feasible point 
$\left( {{\bf{\bar R}},\bar r} \right) = \left( {\beta {{\bf{R}}^*},\beta {r^*}} \right)$ to make ${\rm{tr}}\left( {{{\bf{H}}_S}{\bf{w}}{{\bf{w}}^H}{\bf{H}}_S^H{\bf{R}}} \right) + r\sigma _S^2 = 1$. Obviously, $\left( {{\bf{\bar R}},\bar r} \right)$ meets the constraint (55c) and $\left( {{\bf{\bar R}},\bar r} \right)$ can be proved to provide a smaller optimization value (55a) than that provided from $\left( {{{\bf{R}}^*},{r^*}} \right)$, which is in contradiction with the assumption that $\left( {{{\bf{R}}^*},{r^*}} \right)$ is the optimal solution. Therefore, we can rewrite the constraint (55b) as ${\rm{tr}}\left( {{{\bf{H}}_S}{\bf{w}}{{\bf{w}}^H}{\bf{H}}_S^H{\bf{R}}} \right) + r\sigma _S^2 \ge 1$, which is a convex constraint. Based on the analysis above, (55) can be further expressed as
\begin{align}
	\mathop {\min }\limits_{{\bf{R}}{\underline\succ}0,r \ge 0} \mathop {\max }\limits_{{{\bf{G}}_k} \in {\Upsilon _k}} &\sum\limits_{k = 1}^K {{\rm{tr}}\left( {{{\bf{G}}_k}{\bf{R}}} \right)}  + r\sigma _{}^2\tag{56a}\\
	{{\rm{ s}}{\rm{.t}}{\rm{.}}\quad }&{\rm{tr}}\left( {{{\bf{H}}_S}{\bf{w}}{{\bf{w}}^H}{\bf{H}}_S^H{\bf{R}}} \right) + r\sigma _S^2 \ge 1,\tag{56b}\\
	&(54b)-(54c). \tag{56c}
\end{align}

To simplify the objective function, we introduce the auxiliary variable $t$. Then, (56) can be further written as
\begin{align}
	\mathop {\min }\limits_{{\bf{R}} {\underline \succ}0,r \ge 0}  &t\tag{57a}\\
	{{\rm{ s}}{\rm{.t}}{\rm{.}}\quad }&\mathop {\max }\limits_{{{\bf{G}}_k} \in {\Upsilon _k}} \sum\limits_{k = 1}^K {{\rm{tr}}\left( {{{\bf{G}}_k}{\bf{R}}} \right)}  + r\sigma _{}^2 = t,\tag{57b}\\
	&(56b)-(56c).\tag{57c}
\end{align}Noting that the objective function is to minimize $t$, we can rewrite the non-convex constraint (57b) as a convex constraint $\mathop {\max }\limits_{{{\bf{G}}_k} \in {\Upsilon _k}} \sum\limits_{k = 1}^K {{\rm{tr}}\left( {{{\bf{G}}_k}{\bf{R}}} \right)}  + r\sigma _{}^2 \le t$, which will meet the equality constraint (57b) when the optimal solution of (57) is obtained. Thus, we can formulate the problem (57) as (37).

\section{Proof of Proposition 4}
Firstly, we will prove $\rm{rank}\left( {\bf{R}} \right) = 1 \Rightarrow  \rm{tr}\left( {\bf{R}} \right) - {\lambda _{\max }}\left( {\bf{R}} \right) \le 0$. Since $\rm{rank}\left( {\bf{R}} \right) = 1$, the rank of the column vector group of $\bf{R}$ is 1. Assuming that the first column of $\bf{R}$ is ${\bf{x}}{\rm{ = }}{\left( {{x_{\rm{1}}},{x_2}, \cdots ,{x_N}} \right)^T} \ne 0\left( {{x_{\rm{1}}} \ne 0} \right)$, and the other columns can be represented by $\bf{x}$ linearly, $\bf{R}$ can be represented as ${\bf{R}} = \left( {{y_1}{\bf{x}},{y_2}{\bf{x}}, \cdots ,{y_N}{\bf{x}}} \right) = {\bf{x}}{{\bf{y}}^T}$, where $y_1=1$, ${\bf{y}} = {\left( {{y_1},{y_2}, \cdots ,{y_N}} \right)^T}$. Then, the characteristic polynomials of the matrix $\bf{R}$ can be expressed as 
\setcounter{equation}{57}
\begin{equation}
	\begin{array}{*{20}{l}}
		{\left| {\lambda {{\bf{I}}_N} - {\bf{R}}} \right| = \left| {\begin{array}{*{20}{c}}
					{\lambda  - {x_{\rm{1}}}{y_{\rm{1}}}}&{ - {x_{\rm{1}}}{y_2}}& \cdots &{ - {x_{\rm{1}}}{y_N}}\\
					{ - {x_{\rm{2}}}{y_{\rm{1}}}}&{\lambda  - {x_2}{y_2}}& \cdots &{ - {x_2}{y_N}}\\
					\vdots & \vdots &{}& \vdots \\
					{ - {x_N}{y_{\rm{1}}}}&{ - {x_L}{y_2}}& \cdots &{\lambda  - {x_N}{y_N}}
			\end{array}} \right|}\\
		{\begin{array}{*{20}{c}}
				{{r_i} - \frac{{{x_i}}}{{{x_1}}}{r_1}}\\
				{\overline {\overline {i = 2, \cdots ,N} } }
			\end{array}\left| {\begin{array}{*{20}{c}}
					{\lambda  - {x_{\rm{1}}}{y_{\rm{1}}}}&{ - {x_{\rm{1}}}{y_2}}& \cdots &{ - {x_{\rm{1}}}{y_N}}\\
					{ - \frac{{{x_{\rm{2}}}}}{{{x_{\rm{1}}}}}\lambda }&\lambda & \cdots &{\rm{0}}\\
					\vdots & \vdots &{}& \vdots \\
					{ - \frac{{{x_N}}}{{{x_{\rm{1}}}}}\lambda }&{\rm{0}}& \cdots &\lambda 
			\end{array}} \right|}\\
		{\begin{array}{*{20}{c}}
				{\underline {\underline {{c_1} + \sum\limits_{i = 2}^N {\frac{{{x_i}}}{{{x_1}}}{c_i}} } } }\\
				{}
			\end{array}\left| {\begin{array}{*{20}{c}}
					{\lambda  - \sum\limits_{n = 1}^N {{x_n}{y_n}} }&{ - {x_{\rm{1}}}{y_2}}& \cdots &{ - {x_{\rm{1}}}{y_N}}\\
					0&\lambda & \cdots &{\rm{0}}\\
					\vdots & \vdots &{}& \vdots \\
					0&{\rm{0}}& \cdots &\lambda 
			\end{array}} \right|}\\
		{ = {\lambda ^{N - 1}}\left( {\lambda  - \sum\limits_{n = 1}^N {{x_n}{y_n}} } \right),}
	\end{array}
\end{equation}
where $r_i$ and $c_i$ represent the \emph{i}-th row and column of the matrix ${\lambda {{\bf{I}}_N} - {\bf{R}}}$, respectively.
By letting $\left| {\lambda {{\bf{I}}_N} - {\bf{R}}} \right|$ be zero, we can obtain the eigenvalues of $\bf{R}$ as ${\lambda _1} = {\lambda _2} =  \cdots  = {\lambda _{N - 1}} = 0$, ${\lambda _N} = \sum\limits_{n = 1}^N {{x_n}{y_n}}$. Thus, we have
\begin{equation}
	\begin{split}
		&\rm{tr}\left( {\bf{R}} \right) - {\lambda _{\max }}\left( {\bf{R}} \right) = \sum\limits_{n = 1}^N {{\lambda _n}}  - {\lambda _{\max }}\left( {\bf{R}} \right)\\
		&= {\lambda _N} - {\lambda _{\max }}\left( {\bf{R}} \right) = \left\{ {\begin{array}{*{20}{c}}
				{{\lambda _N},}&{{\lambda _N} < 0}\\
				{0,}&{{\lambda _N} \ge 0}
		\end{array}} \right.
		\le 0.
	\end{split}
\end{equation}

Next, we will prove ${\rm{tr}}\left( {\bf{R}} \right) - {\lambda _{\max }}\left( {\bf{R}} \right) \le 0 \Rightarrow {\rm{rank}}\left( {\bf{R}} \right) = 1$.  Due to the fact that $\rm{tr}\left( {\bf{R}} \right) - {\lambda _{\max }}\left( {\bf{R}} \right) \ge 0$ always holds for any matrix ${\bf{R}}{\underline \succ}0$, $\rm{tr}\left( {\bf{R}} \right) - {\lambda _{\max }}\left( {\bf{R}} \right) \le 0$ can be written as ${\rm{tr}}\left( {\bf{R}} \right) - {\lambda _{\max }}\left( {\bf{R}} \right) = 0$ equivalently. Denote the eigenvalues of $\bf{R}({\bf{R}}{\underline \succ}0)$ as ${\lambda _1},{\lambda _2}, \cdots ,{\lambda _N}\left( {0\le{\lambda _1} \le {\lambda _2} \le  \cdots  \le {\lambda _N}} \right)$, thus, we have
\begin{equation}
	\begin{split}
		&{\rm{tr}}\left( {\bf{R}} \right) - {\lambda _{\max }}\left( {\bf{R}} \right) = \sum\limits_{n = 1}^N {{\lambda _n}}  - {\lambda _{\max }}\left( {\bf{R}} \right) = 0\\
		&\Rightarrow \sum\limits_{n = 1}^N {{\lambda _n}}  = {\lambda _{\max }}\left( {\bf{R}} \right) \Rightarrow {\lambda _N} + \sum\limits_{n = 1}^{N - 1} {{\lambda _n}}  = {\lambda _N}\\
		&\Rightarrow \sum\limits_{n = 1}^{N - 1} {{\lambda _n}}  = 0.
	\end{split}
\end{equation}
Since ${{\lambda _n} \ge 0}$, we can further obtain ${\lambda _1} = {\lambda _2} =  \cdots  = {\lambda _{N - 1}} = 0$. If $\lambda_N=0$, then the positive semi-definite matrix $\bf{R}$ will be a zero matrix, which cannot be an optimal solution. Thus, $\lambda_N$ must be positive, meaning that the positive semi-definite matrix $\bf{R}$ has one and only one non-zero eigenvalue, namely, ${\text{rank}}({\bf{R}})=1$.

\begin{IEEEbiography}[{\includegraphics[width=1in,height=1.25in,clip,keepaspectratio]{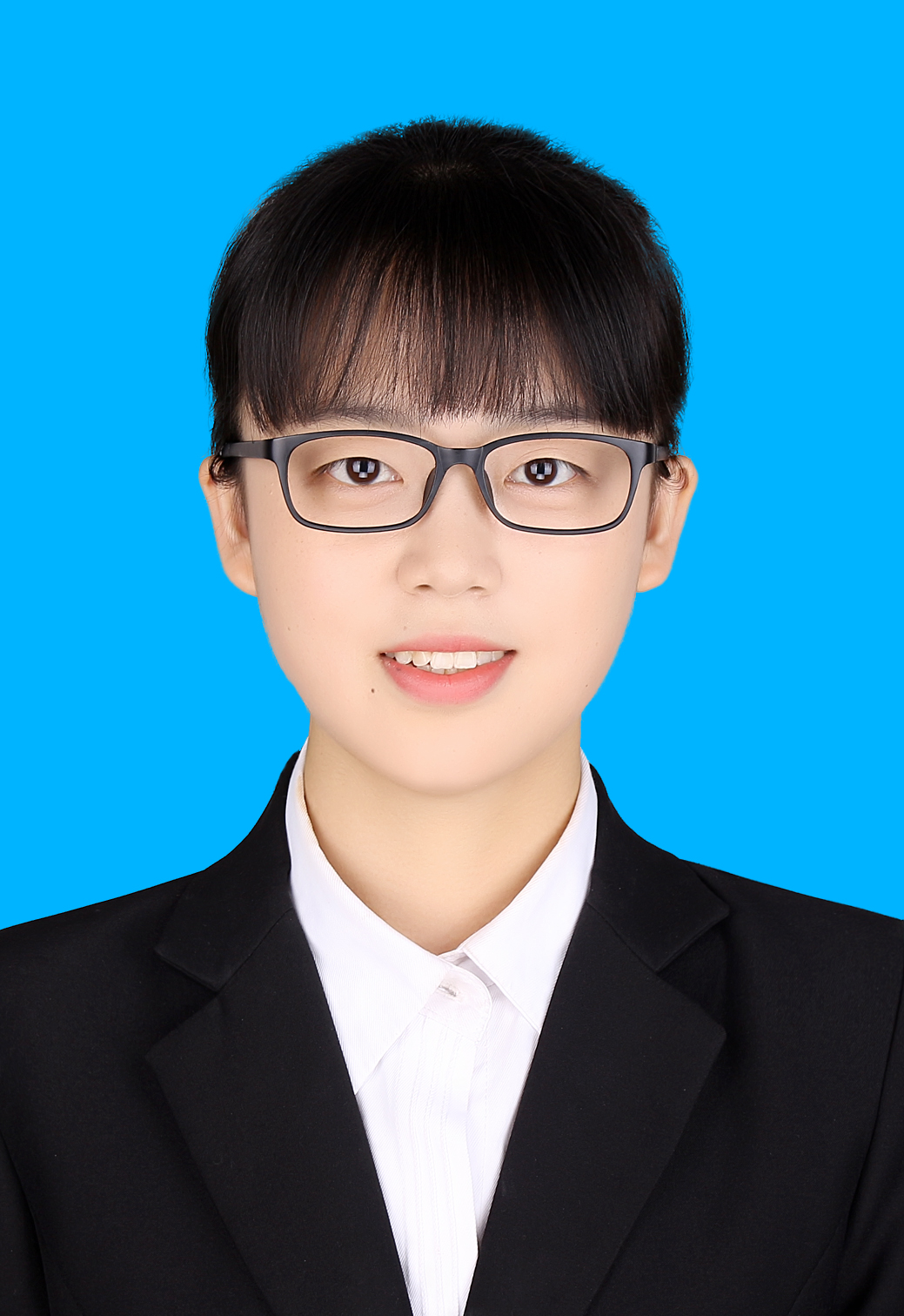}}]{Xuewen Wu}
	(Graduate Student Member, IEEE) received the B.S. degree from College of Computer Science and Technology, Nanjing Tech University, Nanjing, China, in 2017, and the M.S. degree from Nanjing University of Posts and Telecommunications in 2020. She is currently working towards the Ph.D. degree in information and communication engineering from Tongji University. Her research interests include intelligent reflecting surface, cognitive radio, convex optimization, and physical layer security.
\end{IEEEbiography}

\begin{IEEEbiography}[{\includegraphics[width=1in,height=1.25in,clip,keepaspectratio]{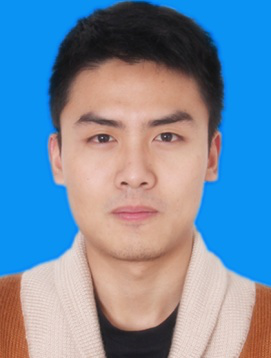}}]{Jingxiao Ma}
	received the B.Eng. degree from the University of Birmingham, U.K., in 2011, the M.Sc. degree in wireless communications from the University of Southampton, U.K., in 2012, and the Ph.D. degree from the University of Sheffield, in 2018. He is currently working with the College of Electronics and Information Engineering, Tongji University, where he is focusing on the areas of relay beamforming, massive MIMO systems, and cognitive radio networks.
\end{IEEEbiography}

\begin{IEEEbiography}[{\includegraphics[width=1in,height=1.25in,clip,keepaspectratio]{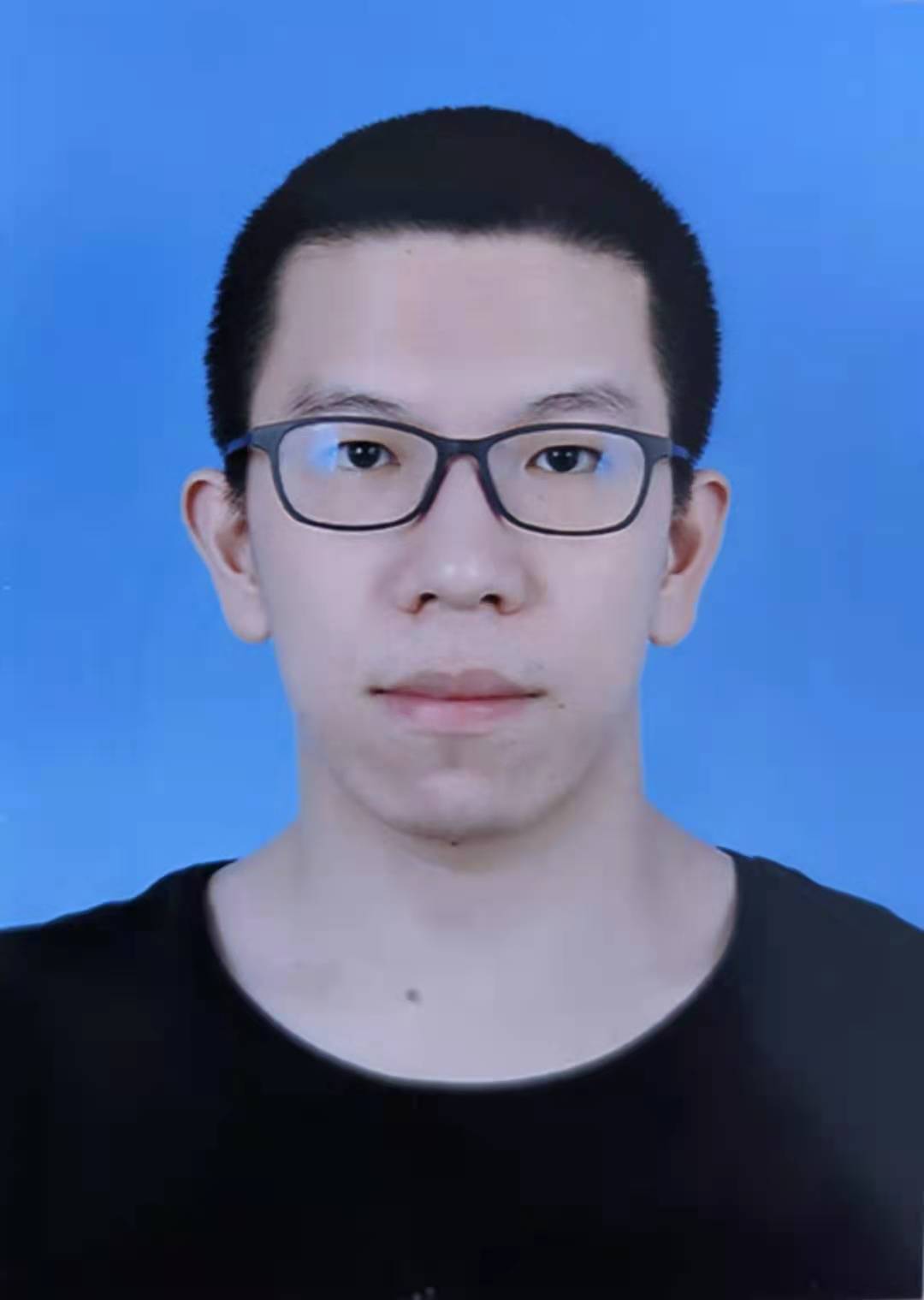}}]{Chenwei Gu}
	received the B.S. degree from Nantong University, Nantong, China, in 2018, and the M.S. degree from Nanjing University of Posts and Telecommunications in 2021. His research interests include wireless communication, convex optimization, and physical layer security.
\end{IEEEbiography}



\begin{IEEEbiography}[{\includegraphics[width=1in,height=1.25in,clip,keepaspectratio]{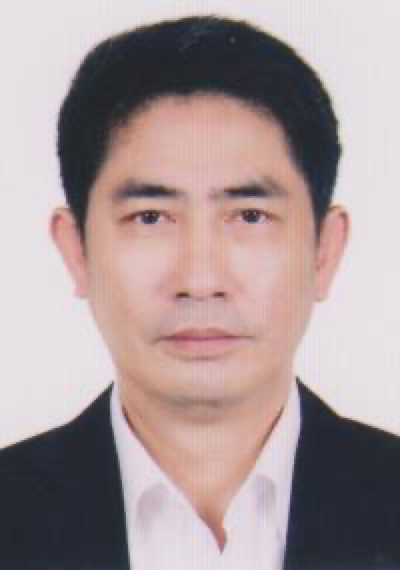}}]{Xiaoping Xue} (Member, IEEE)
	received the B.S. degree in wired communication from Tongji University (formerly Shanghai Railway University), Shanghai, China, in 1987, and the Ph.D. degree in communication and information systems from Beijing Jiaotong University, Beijing, China, in 2009. He is currently a Professor and the Director of the Department of Information and Communication Engineering, Tongji University. His research interests include secure broadband wireless communication theory under high-speed moving conditions, safe computing theory and methods, formal software and security evaluation theory and methods, and security and privacy in vehicular networks.
\end{IEEEbiography}

\begin{IEEEbiography}[{\includegraphics[width=1in,height=1.25in,clip,keepaspectratio]{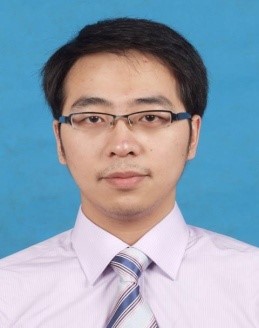}}]{Xin Zeng}
	(Member, IEEE) received the B.Sc. in Communication Engineering from Tongji University, Shanghai, China, in 2009, and the Ph.D in Electronics and Communication from Telecom ParisTech, Paris, France, in 2014. During 2014 and 2018, he was a advanced researcher and standard delegate in Huawei Corporation, focusing on 5G physical layer research and standardization. Since Feb 2018, he has joined Tongji University as an assistant professor in the information and communication department, specialized on wireless communication physical layer design, vehicle networks and applications of artificial intelligence. He has been selected by Shanghai Sailing Program and involved one project from National Natural Science Foundation of China. 
\end{IEEEbiography}
\end{document}